\documentclass[12pt,preprint]{aastex}

\def\msol{\hbox{\kern 0.20em $M_\odot$}}
\def\lsol{\hbox{\kern 0.20em $L_\odot$}}
\def\rsol{\hbox{\kern 0.20em $R_\odot$}}
\def\sr{\hbox{\kern 0.20em sr}}
\def\srmu{\hbox{\kern 0.20em sr$^{-1}$}}

\def\g{\hbox{\kern 0.20em g}}
\def\gmu{\hbox{\kern 0.20em g$^{-1}$}}
\def\kg{\hbox{\kern 0.20em kg}}
\def\pc{\hbox{\kern 0.20em pc}}

\def\mum{\hbox{\kern 0.20em $\mu$m}}
\def\mumd{\hbox{\kern 0.20em $\mu$m$^{-2}$}}
\def\cm{\hbox{\kern 0.20em cm}}
\def\m{\hbox{\kern 0.20em m}}
\def\km{\hbox{\kern 0.20em km}}
\def\nm{\hbox{\kern 0.20em nm}}

\def\s{\hbox{\kern 0.20em s}}
\def\h{\hbox{\kern 0.20em h}}
\def\sec{\hbox{\kern 0.20em sec}}
\def\min{\hbox {\kern 0.20em min}}
\def\smu{\hbox{\kern 0.20em s$^{-1}$}}
\def\smd{\hbox{\kern 0.20em s$^{-2}$}}
\def\an{\hbox{\kern 0.20em an}}
\def\anmu{\hbox{\kern 0.20em an$^{-1}$}}
\def\deg{\hbox{\kern 0.20em $^{\rm o}$}}
\def\yr{\hbox{\kern 0.20em yr}}
\def\yrmu{\hbox{\kern 0.20em yr$^{-1}$}}
\def\Myr{\hbox{\kern 0.20em Myr}}
\def\Mymu{\hbox{\kern 0.20em Myr$^{-1}$}}
\def\K{\hbox{\kern 0.20em K}}
\def\pcmu{\hbox{\kern 0.20em pc$^{-1}$}}
\def\pcmd{\hbox{\kern 0.20em pc$^{-2}$}}
\def\pcmt{\hbox{\kern 0.20em pc$^{-3}$}}
\def\kms{\hbox{\kern 0.20em km\kern 0.20em s$^{-1}$}}
\def\kmpd{\hbox{\kern 0.20em km$^{2}$}}
\def\kpc{\hbox{\kern 0.20em kpc}}
\def\cms{\hbox{\kern 0.20em cm\kern 0.20em s$^{-1}$}}
\def\erg{\hbox{\kern 0.20em erg}}
\def\ergs{\hbox{\kern 0.20em erg}}
\def\cmpd{\hbox{\kern 0.20em cm$^2$}}
\def\cmmd{\hbox{\kern 0.20em cm$^{-2}$}}
\def\cmms{\hbox{\kern 0.20em cm$^{-6}$}}
\def\cmpt{\hbox{\kern 0.20em cm$^3$}}
\def\cmmt{\hbox{\kern 0.20em cm$^{-3}$}}
\def\mpd{\hbox{\kern 0.20em m$^2$}}
\def\mmd{\hbox{\kern 0.20em m$^{-2}$}}
\def\mpt{\hbox{\kern 0.20em m$^3$}}
\def\mmt{\hbox{\kern 0.20em m$^{-3}$}}
\def\mujy{\hbox{\kern 0.20em $\mu$Jy}}
\def\mjy{\hbox{\kern 0.20em mJy}}
\def\Mj{\hbox{\kern 0.20em MJy}}
\def\jy{\hbox{\kern 0.20em Jy}}
\def\ghz{\hbox{\kern 0.20em GHz}}
\def\srmd{\hbox{\kern 0.20em sr$^{-1}$}}
\def \kms{km~$\rm{s}^{-1}$}

\def \mum{$\mu$m}

\def\G{\hbox{\kern 0.20em G}}

\def\h13cop{\hbox{H$^{13}$CO$^{+}$}}

\def\S+{\hbox{S{\small II}}}

\slugcomment{Submitted to ApJ on July 8th 2009}

\shorttitle{Dust Properties of NGC~147 and NGC~185}
\shortauthors{Marleau et al.}

\begin{document}

\newcommand{\jfourteen}{\hbox{$J=14\rightarrow 13$}}
\title{Dust Abundance and Properties in the Nearby Dwarf Galaxies NGC~147 and NGC~185}

\author{Francine R.\ Marleau\altaffilmark{1}, 
Alberto Noriega-Crespo\altaffilmark{2}, 
and 
Karl A.\ Misselt\altaffilmark{3}}

\altaffiltext{1}{Department of Astronomy and Astrophysics, University
of Toronto, 50 Saint George Street, Toronto, ON M5S 3H4, Canada}

\altaffiltext{2}{Spitzer Science Center, California Institute of
Technology, CA 91125 USA}

\altaffiltext{3}{Steward Observatory, University of Arizona, Tucson AZ
85721 USA}

\begin{abstract}
We present new mid- to far-infrared images of the two dwarf compact
elliptical galaxies that are satellites of M31, NGC~185 and NGC~147,
obtained with the Spitzer Space Telescope. Spitzer's high sensitivity
and spatial resolution enable us for the first time to look directly
into the detailed spatial structure and properties of the dust in
these systems. The images of NGC~185 at 8 and 24 micron display a
mixed morphology characterized by a shell-like diffuse emission region
surrounding a central concentration of more intense infrared
emission. The lower resolution images at longer wavelengths show the
same spatial distribution within the central 50\arcsec\ but beyond
this radius, the 160\mum\ emission is more extended than that at 24
and 70\mum. On the other hand, the dwarf galaxy NGC 147 located only a
small distance away from NGC~185 shows no significant infrared
emission beyond 24 micron and therefore its diffuse infrared emission
is mainly stellar in origin. For NGC~185, the derived dust mass based
on the best fit to the spectral energy distribution is $1.9 \times
10^3 \rm{M}_\odot$, implying a gas mass of $3.0 \times 10^5
\rm{M}_\odot$. These values are in agreement with those previously
estimated from infrared as well as CO and HI observations and are
consistent with the predicted mass return from dying stars based on
the last burst of star formation $1 \times 10^9$~yr ago. Based on the
70 to 160\mum\ flux density ratio, we estimate a temperature for the
dust of $\sim$~17K. For NGC~147, we obtain an upper limit for the dust
mass of $4.5 \times 10^{2} \rm{M}_\odot$ at 160\mum\ (assuming a
temperature of $\sim$~20K), a value consistent with the previous upper
limit derived using ISO observations of this galaxy. In the case of
NGC~185, we also present full $5-38$\mum\ low-resolution (R$\sim$100)
spectra of the main emission regions. The IRS spectra of NGC~185 show
strong Polycyclic Aromatic Hydrocarbons (PAH) emission, deep silicate
absorption features and H$_2$ pure rotational line ratios consistent
with having the dust and molecular gas inside the dust cloud being
impinged by the far-ultraviolet radiation field of a relatively young
stellar population. Therefore, based on its infrared spectral
properties, NGC~185 shows signatures of recent star formation (a few
$\times 10^8$ years ago), although its current star formation rate is
quite low.
\end{abstract}


\lefthead{Marleau et al.}

\righthead{Spitzer Observations of NGC~147 and NGC~185}

\section{Introduction}

It was by resolving into stars the two dwarf elliptical galaxies
NGC~147 and NGC~185, along with their formidable large companion the
Andromeda galaxy, that Baade refined his concept of two stellar
populations in the 1940s. Of NGC~147, Baade (1944) wrote ``In contrast
with the uncertain information provided by the blue-sensitive plates,
the 4-hour red exposure of NGC~147 is truly revealing. It shows that
the nebula is a large star cloud, ellipsoidal in structure, and of a
density gradient so low that even the central region is fully
resolved''. He later on remarked that NGC~147 is an example of a
dust-free galaxy containing only stars of Population II (Baade
1951). As part of the same work, Baade's notes on NGC~185 revealed
``NGC~185, considerably brighter than NGC~147, has been classified by
Hubble as Ep, the peculiarity being the abnormally slow increase in
intensity toward the center. It is intermediate in this respect to
NGC~205, similarly classified by Hubble as Ep, and NGC~147. NGC~185 is
one of the few elliptical nebulae in which patches of obscuring
material are conspicuous. Two such dark clouds are near the center of
NGC~185''. Based on his observation of bright young blue stars in
NGC~185, he also noted the apparent relationship between these
Population I (metal-rich) stars and the presence of dust.

Based on our current understanding of stellar evolution, we now know
that it is the dying stars that inject gas and metal-enriched dust
back into the interstellar medium (ISM) (Faber \& Gallagher 1976). The
presence of planetary nebulae in NGC~185 (Ford, Jenner \& Epps 1973),
for example, confirms that mass loss is occurring in this dwarf
elliptical galaxy. From this enriched material is formed a new
population of luminous hot and young metal-rich stars. Indeed, the ISM
in NGC~185, as seen in the form of atomic (HI) and molecular gas (CO),
appears spatially concentrated near the present-day star-forming
region (Young 2001; Welch, Mitchell \& Yi 1996). However, whereas
NGC~185 contains gas, dust and young stars with ages of about 100~Myr
(in the central 150~pc by 90~pc region; Martinez-Delgado, Aparicio, \&
Gallart 1999) to 400~Myr (Butler \& Martinez-Delgado 2005), it is
unknown why NGC~147 appears to be presently dust- and gas-free (Young
\& Lo 1997; Sage, Welch, \& Mitchell 1998) and had its most recent
star forming episode $\lesssim$~1~Gyr ago (Han et al.\ 1997). Either
an efficient removal mechanism of some kind is at play (external
sweeping, formation of condensed objects, gas ejection by violent
nuclear activity, or sweeping by a hot galactic wind) or NGC~147 is at
a completely different evolutionary stage than NGC~185. Although there
are some claims that these two dwarf galaxies form a stable binary
system (van den Bergh 1998), the latter explanation implies that these
two galaxies did not form at the same time and/or in the same
environment despite their physical proximity.

In this paper, we explore the nature of these two systems from a
completely new direction, using a different wavelength regime. We
present new mid- to far-infrared images of NGC~147 and NGC~185
obtained with the {\it Spitzer Space Telescope} (hereafter {\it
Spitzer}). The higher sensitivity and spatial resolution of {\it
Spitzer} over previous infrared observatories enable us for the first
time to measure {\it directly} the detailed structure and composition
of the dust in these two dwarf galaxies.  The paper is organized as
follows. In Section~2, we describe the {\it Spitzer} observations and
the data reduction. The {\it Spitzer} images are presented in
Section~2, followed by a description of the mid- and far-infrared
emission morphology and a comparison with other tracers of the ISM in
Section~3. Foreground point source contamination is discussed in
Section~4. Dust mass measurements and gas mass estimates are derived
in Section~6 based on the spectral energy distribution of the dwarf
galaxies presented in Section~5. In Section~7, we present the results
of the analysis of the first set of {\it Spitzer} spectroscopic
observations of one of M31's dwarf galaxies. We discuss the spectral
line measurements and spectral properties of NGC~185's dust clouds. We
conclude our paper by presenting a summary of our observations and
results in Section~8.

\section{Observations and Data Reduction}

\subsection{IRAC and MIPS Images}

The mid-infrared imaging observations of the two dwarf galaxies
NGC~147 and NGC~185, shown in the optical in Figure~1, were obtained
with IRAC, the infrared camera (Fazio et al.\ 2004) on board {\it
Spitzer} (Werner et al.\ 2004) on 2005 July 25 and 2005 August 20,
respectively, as part of the GO2 program to study in detail the dust
in M31's four elliptical companions (PI: Marleau, Program ID: 20173).
A region 5\arcmin\ in size was mapped in each of the four IRAC
channels using a total integration time of 84 seconds per sky
position. The 24, 70 and 160\mum\ MIPS (the Multiband Imaging
Photometer for {\it Spitzer}; Rieke et al.\ 2004) photometric
observations of both galaxies were completed on 2006 February 16 (24
\& 70\mum) and on 2005 September 04 (160\mum) with an integration time
per pixel of 606, 440 and 63 seconds, respectively.

The images obtained with channels 1 and 2 (at 3.6 and 4.5\mum) were
used mostly to determine the stellar contribution to the overall
spectral energy distribution of the galaxy. Channels 3 and 4 (at 5.8
and 8\mum) contain the most information about the dust properties;
channel 4 in particular, with a passband centered at 7.9\mum\ includes
some of the strongest Polycyclic Aromatic Hydrocarbons (PAH) emission
features known (e.g.\ bands at 7.7 and 8.6\mum); these are considered
some of the best tracers of very small particles (see e.g.\ Li \&
Draine 2001, Draine \& Li 2001 and references therein).

The IRAC Basic Calibrated Data (BCD) produced by the SSC pipeline
version S14.0.0 (2006 May) were combined with the SSC MOPEX software
to produce final mosaics with pixel size 1.2\arcsec\ (all channels)
and spatial resolutions of less than 2\arcsec\ FWHM. The MIPS BCDs
used in this paper were generated with the SSC pipeline version
S13.2.0 (2005 November - 2006 January). These versions of the SSC
software are quite stable and adequate for processing NGC185 and
NGC147 since the brightness of both systems is outside the regime
affected by non-linearity, latencies or muxbleed effects.

The MIPS 24\mum\ data suffer from the ``First-frame Effect,'' where
the first, and often the second and third, frame of every commanded
sequence of observations have a shorter exposure time and are
depressed in response by as much as $10-15$\%. Therefore, based on the
MIPS IST recommendation, the first two BCD frames of every commanded
sequence of observations were discarded. Sometimes the pipeline
flat-fielding does not completely remove background gradient across
the MIPS 24\mum\ frames. Therefore, we performed an extra
flat-fielding of our BCDs by ``self-calibrating'' each BCD, i.e.\
flattening (dividing) each BCD by the normalized median BCD image. As
the dither positions between BCD images have the emission from the
galaxy overlapping, the galaxy was masked before the median was
calculated. We created our own median by masking the galaxy and
replacing the masked pixels by the median of the non-masked pixels in
each BCD image. Further processing of the BCDs consisted of removing
the zodiacal light component of each frame by subtracting the estimate
taken from the header keyword ZODY\_EST (the median value across the
stack of BCDs was 24.17~MJy/sr for NGC~147 and 24.36~MJy/sr for
NGC~185). To perform background correction of our BCDs we used
``overlap.pl'' in MOPEX. This levels the backgrounds to ensure that
the final mosaic is not ``patchy''. Finally, once all these
corrections were applied, the processed BCDs were re-mosaicked using
``mosaic.pl'' in MOPEX.

The MIPS 70\mum\ data contain the ``stim'' BCD frames. Therefore,
these stim BCD images were removed from the stack of BCDs. In total,
90 BCDs were removed out of the 468 BCD frames for each galaxy
(leaving 378 BCD frames each). Also, variation across columns are seen
at 70\mum\ due to readout. These variations were removed by
subtracting the median of the values along each column for every BCD
(masking the NaNs and the galaxy), a procedure referred to as ``column
filtering''. A median value of the whole image (of pixels that were
not masked, i.e.\ NaNs or belonging to the galaxy) was added back to
retain the background flux value. As for MIPS 24\mum, further
processing of the BCDs consisted of removing the zodiacal light
component of each frame by subtracting the estimate taken from the
header keyword ZODY\_EST (the median value across the stack of BCDs
was 6.07~MJy/sr for NGC~147 and 6.14~MJy/sr for NGC~185). Given that
the coverage of the final mosaic is non-uniform (lower coverage
regions in the bottom left of the galaxy), we performed a background
correction of our BCDs by using ``overlap.pl'' in MOPEX. Finally, once
all these corrections were applied, the processed BCDs were
re-mosaicked using ``mosaic.pl'' in MOPEX.

As for the MIPS 70\mum\ data, the MIPS 160\mum\ data contain as well
the ``stim'' BCD frames. Therefore, these stim BCD images were also
removed from the stack of BCDs. In total, 72 BCDs were removed out of
the 468 BCD frames for each galaxy (leaving 396 BCD frames each).
Further processing of the BCDs consisted of removing the zodiacal
light component of each frame by subtracting the estimate taken from
the header keyword ZODY\_EST (the median value across the stack of
BCDs was 0.93~MJy/sr for NGC~147 and 0.94~MJy/sr for NGC~185). No
overlap correction was applied since the array is small and has a
gap. Finally, once all these corrections were applied, the processed
BCDs were re-mosaicked using ``mosaic.pl'' in MOPEX.

The final 24, 70 and 160\mum\ mosaics were produced with square
pixels, 1.225, 4.0 and 8.0\arcsec\ in size and had a final spatial
resolution of 6, 18 and 40\arcsec\ FWHM, respectively (Rieke et al.\
2004). The astrometry was checked by comparing 2MASS point sources on
IRAC 8\mum\ images and then comparing the 8\mum\ point sources with
24\mum. The astrometry was found accurate to within the astrometric
accuracy of 2MASS (0.2\arcsec) and no correction was applied. The
final IRAC and MIPS mosaics of NGC~147 and NGC~185 are shown in Figure
2 \& 3 and 4 \& 5, respectively.

\subsection{IRS Spectra}

Observations of NGC~185 at three different dust cloud location visible
in the optical image of the galaxy were performed using the Infrared
Spectrograph (IRS; Houck et al.\ 2004) on board {\it Spitzer} on 2006
January 29-30 (see Table~1 and Figure~6). In addition, a separate
observation of the sky (see Table~1, ``NGC185IRSSky''), located
6\arcmin\ away from the center of NGC~185, was obtained for accurate
background subtraction. All four modules, Short-Low 2 (SL2;
$5.2-8.7$\mum) and 1 (SL1; $7.4-14.5$\mum), Long-Low 2 (LL2;
$14.0-21.3$\mum) and 1 (LL1; $19.5-38.0$\mum), were used, to obtain
full $5-35$\mum\ low-resolution (R$\sim$100) spectra. There were two
nodding positions per observation and a number of 10(SL)/20(LL) cycles
of 60(SL)/30(LL) seconds ramp duration were used at each nod position
giving a total on-source integration time of 1219 (SL) and 1258 (LL)
seconds per pixel.

The IRS data used in this paper were generated by the SSC pipeline
version S13.2.0 (2006 January). The data reduction consisted of the
following steps. First, the four nod positions of the sky observations
in each channel were combined (by computing the median value) to
create a ``super sky''. The uncertainty image associated with the
super sky was also created. Second, the sky was subtracted from each
``\_coa2d.fits'' spectral image of the dust clouds. The uncertainty
images associated with the background subtracted image was also
computed. Third, we used the IDL interactive tool
IRSCLEAN\_MASK\footnote{http://ssc.spitzer.caltech.edu/archanaly/contributed/irsclean/IRSCLEAN\_MASK.html}
to create a mask of rogue pixels associated with the background
subtracted spectral images and to clean those images prior to spectral
extraction. Finally, spectral extraction was done using the Spitzer
IRS Custom Extraction software
SPICE\footnote{http://ssc.spitzer.caltech.edu/postbcd/spice.html}
using the ``extended source with regular extraction'' template. The
first step of the extraction was done using the module called
``profile'' which computes and plots the mean spatial flux
profile. The second step was done using the module called ``ridge''
which finds the peak in the spatial profile. For this step, we
selected the option ``manual'' to ensure that the extraction was
always done at the same location and for the same source (since
sometimes there were more than one peak) at the two nod positions. The
extraction was done with the ``extract'' module. For this, we did not
use the full width extraction (default 28 pixels) but we specified an
extraction of width 7 pixels. This is because the full width
extraction introduces spurious lines at the very edge of the
extraction width where the spectral image is not as clean and the
narrower extraction also produces a better signal-to-noise
spectrum. The extraction spectra were flux calibrated using the module
called ``tune'' which produces two output spectra,
``aploss\_spect.tbl'' and ``extsrc.tbl''. The former is the spectrum
derived assuming a point source extracted with the full aperture and
the latter is the spectrum ``aploss\_spect.tbl'' multiplied by the
slit loss correction. The slit loss correction assumes that at a given
wavelength the source is infinite in extent and has a constant surface
brightness. As we did not use a full width extraction, we did our own
flux calibration of the ``extsrc.tbl'' spectra (see below). The
spectral extraction procedure using SPICE is illustrated in
Figure~7. After extraction was performed, we averaged together the
spectra at the two nod positions for each of the SL2, SL1, LL2 and LL1
wavelength coverage and trimmed the noisy edges.

The 1st and 2nd order of the SL and LL spectra were found to be in
good agreement. However, this procedure left a mismatch between the SL
and the LL regions of the spectra. We therefore scaled the SL to match
the LL (using a multiplicative factor; IRS Data Handbook Version 3.1,
2007). To properly flux calibrate our spectra, as we did not use a
full width extraction, we used our flux density measurements obtained
from the 24\mum\ images. As our extraction width remained exactly the
same for all IRS target positions, we extracted the photometry for
``NGC185IRSNorth'' (brightest cloud) and applied the same additive
offset to the other two spectra, keeping their relative brightness the
same. The aperture for the photometry was selected to be centered on
the target position for ``NGC185IRSNorth'', with an area of
0.106\arcmin$^2$, matching the IRS extraction area: the slit width
(10.7\arcsec) multiplied by the width of extraction (7~pix $\times$
5.1\arcsec/pix = 35.7\arcsec). For ``NGC185IRSNorth'', after
background subtraction, we obtained a flux density measurement of
3.8~$\pm$~0.4~mJy. The ``NGC185IRSNorth'' spectrum was therefore flux
calibrated using this measurement and the other spectra were offset
accordingly.

\section{Morphology of Infrared Emission}

\subsection{NGC~147}

Figure~8 displays a three-color image, as seen by IRAC at 3.6, 5.8 and
8\mum, of NGC~147. At the depth of our observations, we are unable to
detect significant dust emission beyond 24\mum. There is a small
contribution at shorter wavelength (at 24, 8 and 5.8\mum), however,
the dominant emission at these wavelengths, as seen in Figure~8, is
stellar in origin. The stellar emission extends to about 4.2\arcmin\
(823~pc) along the major axis of the galaxy and 2.2\arcmin\ (431~pc)
along the minor axis. At a distance of 675~$\pm$~27~kpc (McConnachie
et al.\ 2005), 1.0\arcmin\ corresponds to 196~pc. The two prominent
extended sources easily detected at 70\mum\ (Figure~3) can be easily
identified as background galaxies based on their morphology and color
(see Figure~8).

The lack of significant dust emission in this galaxy is consistent
with observations done at 20~cm with the Very Large Array (VLA; Young
\& Lo 1997) looking for HI emission and at 115~GHz with the NRAO 12 m
telescope at Kitt Peak (Sage, Welch, \& Mitchell 1998) searching for
the CO J=1-0 line emission. Both HI and CO remain undetected in this
galaxy.

\subsection{NGC~185}

Figure~9 displays an IRAC three-color image of NGC~185 where the dust
clouds at 5.8 and 8\mum\ stand out within the stellar background
detected by the shorter wavelength bands. 

As seen from these images, especially the highest resolution image
available at 8\mum, the diffuse dust emission from NGC~185 has a mixed
morphology characterized by a shell-like emission region extending
from the south to the east of the galaxy center surrounding a central
region of more concentrated emission. The mechanism responsible for
such a morphology is not known but, as can be inferred from Figure~9,
it is likely the result of the recent star forming activity detected
in this galaxy. The morphology of the dust emission at longer
wavelengths is similar to that seen at shorter wavelengths in the
central region. Even at 70\mum, where the spatial resolution is
18\arcsec\ FWHM, we can still distinguish the shell-like structure and
central emission region. At 160\mum\ (40\arcsec\ FWHM), the spatial
structure of the emission is unresolved.

The peak of the emission is concentrated at the center in a region
$\sim$ 30\arcsec\ in diameter. At the adopted distance of
616~$\pm$~26~kpc (McConnachie et al.\ 2005), 1.0\arcmin\ corresponds
to 179~pc. Therefore, an angular size of 30\arcsec\ corresponds to a
physical size of 89~pc. However, the diffuse dust emission is clearly
seen to extend to a much larger distance, covering a region
approximately 2.5\arcmin\ diameter in size (or 447~pc).

The ISM in NGC~185 is concentrated near the present-day star-forming
region, although the atomic and molecular components are not spatially
coincident (CO, Welch, Mitchell \& Yi 1996; HI, Young 2001). As shown
in Figure~10, the location of the main peak of each of the CO and HI
emission does not match exactly the location of the peak of the dust
emission, although they both are coming from the same dust cloud
(``NGC185North'', see Table~4 and Figure~16). The morphology and
kinematic properties of the ISM are consistent with it originating
from stars internal to NGC~185 (Young \& Lo 1997; Martinez-Delgado,
Aparicio \& Gallart 1999).

Figure~11 shows the MIPS three-color image of NGC~185. In order to
understand the distribution of the cold dust in NGC~185 as traced by
the 160\mum\ light, we have performed some radial cuts along the dust
clouds symmetry axis and compared them with the 24 and 70\mum\
emission that trace other dust temperature and size (small grains)
components. This comparison requires that we match the 24 and 70\mum\
beams to that of the 160\mum\ array (FWHM~$\sim$~40\arcsec). To
accomplish this, we convolved the 24 and 70\mum\ images with their
corresponding kernels selecting a PSF representative of 100~K black
body emission (see e.g.\ Gordon et al.\ 2008). Figure~12 shows one of
the surface brightness cuts (at a position angle of 52\deg, i.e.\
along the major-axis of the infrared diffuse emission) normalized and
shifted so that the three bands start with a common baseline near
zero. The radial distribution of the three MIPS bands is similar
within the central 50\arcsec\ but beyond this radius, the 160\mum\
emission is more extended than that at 24 and 70\mum, a trend observed
in other galactic systems (e.g.\ Engelbracht et al.\ 2004; Hinz et
al.\ 2006).

\section{Resolved Stars and Foreground Contamination}

Point sources were extracted in the central 5\arcmin\ $\times$
5\arcmin\ region of both dwarf galaxies in the 5.8, 8.0 and 24\mum\
images using the source extraction software StarFinder (Diolaiti et
al.\ 2000). {\it Spitzer} $[8]-[24]$ and $[5.8]-[8]$ colors were
derived for these point sources and compared to the colors of a sample
of Galactic AGB and K Giant star templates in the mid-infrared
observed with the Short Wavelength Spectrometer (SWS, covering a
wavelength range of $2.4-45.2$\mum) on board of the Infrared Space
Observatory (ISO) in order to identify the type of resolved stars
present in NGC~147 and NGC~185 and to estimate the amount of
foreground contamination from our Galaxy in their direction (the
contamination from M31 is considered negligible at this large distance
from their large companion). For each galaxy, we detected 34 point
sources with emission in all of the 5.8, 8 and 24\mum\ bands, allowing
us to compute their colors. The photometry of the point sources in
NGC~147 and NGC~185 is given in Table~2 and 3, respectively.

The flux densities within each of the Spitzer's imager/photometer
bandpass of a sample of Galactic AGB and K Giant star templates were
determined using ISAP (ISO Spectral Analysis Package Version
2.0)\footnote{ISAP is available at
http://www.ipac.caltech.edu/iso/isap/isap.html}. As illustrated in
Figure~13 and 14, the majority of point sources in the central region
of NGC~147 and NGC~185 have colors similar to those of AGB
stars. However, a handful of K Giant stars are clearly identifiable in
each field due to their much ``bluer'' $[8]-[24]$ and $[5.8]-[8]$
colors. For the K Giant classification in Table~2 and 3, we used the
selection criteria of $[8]-[24]<0.8$ and $[5.8]-[8]<0.3$.

The foreground contamination at 5.8, 8, and 24\mum\ from Galactic
stars in the ``Total'' emission region (as defined in Table~4 and seen
in Figure~15 and 16) of each dwarf galaxy was estimated using a
control field located 7\arcmin\ away from each dwarf galaxy. For both
dwarf galaxy, we estimated a Galactic contamination of about 20\%,
with the remaining 80\% of the infrared point sources belonging to the
galaxies themselves. This implies that we detected and resolved
$\sim$~28~dusty AGBs in NGC~147 and 185, respectively.

\section{Spectral Energy Distributions}

We measured the IRAC 3.6, 4.5, 5.8, \& 8\mum\ and the MIPS 24, 70, \&
160\mum\ flux densities on the images in the spatially resolved
regions common at all wavelengths, as depicted in Figure~15 and 16 and
listed in Table~5 and 6. In the case of NGC~147, we were unable to
detect significant dust emission at 70 and 160\mum. The values
reported in Table~5 at these wavelengths are the 1$\sigma$ sensitivity
upper limits measured from the standard deviation maps.
The flux uncertainty for each region
includes the combined effect of the systematic errors due to the post
processing of the images, plus that of the flux density calibration of
the instruments (Fazio et al.\ 2004; Rieke et al.\ 2004). The flux
uncertainty, therefore, was estimated by measuring the residual flux
density away from the galaxy in the background subtracted images and
adding an absolute flux calibration uncertainty of 10\% for the IRAC
and MIPS 24\mum\ measurements, and of 20\% for the MIPS 70 \& 160\mum\
flux densities. The IRAC flux densities are corrected (multiplied) by
the effective aperture correction factor of 0.944 (3.6\mum), 0.937
(4.5\mum), 0.772 (5.8\mum), and 0.737 (8\mum) (Reach et al.\
2005). The corrected flux densities and uncertainties are given in
Table~5 and 6.

The stellar photospheric component of the diffuse emission coming from
each galaxy itself must be removed before the analysis of the dust
component. We conservatively assume that, in addition to the 2MASS J,
H, and K bands, all the light in the 3.6 and 4.5 IRAC bands is
photospheric in origin. The J, H, K, IRAC 3.6 and 4.5 filter response
functions were convolved with a PEGASE stellar evolutionary synthesis
(SES) models (Fioc \& Rocca-Volmerange 1997) with an age of 1~Gyr and
a metallicity of [Fe/H]~=~-0.7. The SES models were computed at a
metallicity of -0.7 since it is the closest metallicity point in our
SES model grids to the measured metallicity of both galaxies
([Fe/H]~=~-1.2 and -1.1 for NGC~185 and NGC~147, respectively; Mateo
1998; McConnachie et al.\ 2005). However, the broad band dust SED
fitting is not sensitive to the chosen metallicity and model runs at
the two neighboring grid points in our SES models yielded identical
results. There is also no diagnostic power in the age assumed; the
PEGASE models were used here only to estimate the stellar contribution
to the Spitzer bands, as simple power laws cannot reproduce the J, H,
K colors. Scalings were derived for each band individually and
averaged together to get the final scaling for the photospheric
component. In all cases, the J through 4.5\mum\ flux densities are
consistent with being completely stellar in origin. To estimate the
photospheric contribution to the flux density at longer wavelengths,
we convolved the relevant IRAC and MIPS bandpasses with the PEGASE
model and subtracted the result from the observed flux density to
yield an estimate for the total dust flux density. For the relative
contributions of the diffuse photospheric emission to the observed
flux densities for NGC~185, see Table~7. The age we use for the PEGASE
models does not significantly alter the dust fits; indeed, PEGASE
models between 0.001$-$10~Gyr result in very similar estimates for the
stellar contamination in the Spitzer bands, especially for $\lambda
\ge 8$\mum.

\section{Mass of Dust and Gas using SED Fitting}

We can determine dust masses by fitting the total dust emission
SED. The SED was measured between 5.8 and 160\mum, using IRAC and MIPS
data. We measure the flux densities in the spatially resolved regions
common to all wavelengths, as depicted in Figure~15 and 16 and listed
in Table~5 and 6.

Modeling the dust emission in the dwarf galaxies requires the
specification of both a dust model and a source of heating in the form
of the radiation field the grains are exposed to. Grains exposed to
the radiation field will absorb energy and emit according to their
size and composition. Larger grains will tend to achieve an
equilibrium temperature while smaller grains will undergo stochastic
temperature fluctuations (Draine \& Li 2001; Misselt et al.\
2001). The temperature a given grain achieves (and whether it reaches
equilibrium) is determined by the shape and intensity of the radiation
field. Once the emission spectrum of each grain in the dust model has
been computed, the total dust emission spectrum can be computed and
scaled to fit the observations, where the scaling depends only on the
dust mass and the distance to the source. Clearly, specifying the dust
model and radiation field introduces a large number of free parameters
to the fitting procedure: dust size distributions, dust composition,
radiation field shape and intensity (Draine \& Li 2007).  To reduce
the number of free parameters in our fit, we fix the dust model by
adopting the BARE-GR-S model of Zubko et al.\ (2004) while to
determine the radiation field, we adopt the PEGASE SES models (Fioc \&
Rocca-Volmerange 1997). The Zubko et al.\ (2004) dust model consists
of three components, polycyclic aromatic hydrocarbons (PAH), graphitic
grains, and silicate grains. With these assumptions, we have four free
parameters in our fit: 1) the shape of the radiation field; 2) the
intensity of the radiation field; 3) the total dust mass; 4) the
distance to the galaxy. We fix the distances at 675 and 616~Mpc for
NGC~147 and NGC~185, respectively (McConnachie et al.\ 2005). The
shape of the radiation field is set by the age of the PEGASE SES
model. The intensity of the field is set by first normalizing the SES
to have the same energy content as the local interstellar radiation
field as defined by Mathis, Mezger \& Panagia (1983). The intensity is
then parametrized by a dimensionless scaling factor applied to the
normalized SES model.

With the above assumptions, the fitting proceeds as follows. Since
there is no analytic relation between the radiation field and the dust
emission spectrum, we take the approach of computing a large grid of
possible galaxy SEDs and select the best fit (in the $\chi^2$ sense)
to the observed SED from the set of model SEDs. To compute the grid of
model SEDs, we compute 60 PEGASE SES models with ages ranging from
0-10~Gyr. For each age, the model stellar SED is normalized to the
Mathis, Mezger \& Panagia (1983) local radiation field and 200
dimensionless scalings between 0.1 and 2000 are applied. For each of
the resulting 12000 model radiation fields, the dust emission SED is
computed from the Zubko et al.\ (2004) BARE-GR-S model following the
prescription of e.g.\ Misselt et al.\ (2001), yielding the dust
emission per unit dust mass. This dust SED is then fitted to the
observed SED with the only free parameter being the dust mass. Fixing
the dust model as we have done results in a fixed relationship between
the relative masses in each dust component (PAH, graphite, and
silicates). Ranges for each parameter were measured by constructing
68\% confidence intervals in the three dimensional parameter space.

\subsection{NGC~147}

Since only upper limits are available at 70 and 160\mum\ (see Table~5
and Figure~17), we have almost no constraints on the dust mass for
NGC~147 from SED fitting. The fitting procedure outlined above results
in two classes of solutions that are able to fit the 8 and 24\mum\
observations while not violating the 70 and 160\mum\ upper limits. The
first class is characterized by a small mass
($\lesssim$~few~$\rm{M}_\odot$) of warm dust, whereas the second class
is characterized by a large mass ($\gtrsim 10^{7} \rm{M}_\odot$) of
cold dust. The low mass class of solutions satisfy the upper limit
requirements by keeping the peak of the dust emission shortward of
$\sim$40\mum\ whereas the cold, massive solutions shift the peak of
dust emission well longward of 160\mum. However, the stellar component
can be well fitted by the PEGASE model and there is no indication of
an infrared excess at 8\mum\ in this galaxy.

The current data and our fitting procedure provide little diagnostic
power with respect to the properties of the ISM in NGC~147. However,
some general observations can be made. The cold, massive solutions are
unrealistic given the upper limit of the HI mass of $3 \times 10^{3}
\rm{M}_\odot$ (Young \& Lo 1997). On the other hand, the warm low mass
solutions with dust masses between a few tenths to $\lesssim 1
\rm{M}_\odot$ are consistent with the upper limit of $4.1 \times 10^2
\rm{M}_\odot$ derived by Temi et al.\ (2004) (see
below). Additionally, taking a gas-to-dust mass ratio of 160
appropriate for the Zubko et al.\ (2004) BARE-GR-S dust model, the gas
masses implied by the warm, low mass solutions are of the order of
$10^{2}\rm{M}_\odot$, a factor of $10^2$ less than the upper limit of
$9.9 \times 10^3 \rm{M}_\odot$ derived by Sage, Welch \& Mitchell
(1998). On the other hand, neither the Sage, Welch \& Mitchell (1998)
upper limit nor the gas masses implied by our low mass class of
solutions, is consistent with the estimated return of gas from evolved
stars to the ISM of $5.5 \times 10^5 \rm{M}_\odot$ calculated using
the prescription of Faber \& Gallagher (1976) for a time of $5 \times
10^8$~yr.

Similarly to Temi et al.\ (2004), we can also compute an upper limit
on the dust mass in this galaxy from the upper limit on the flux
density at 160\mum\ using the standard relation (Hildebrand 1983):

\begin{equation}
M_d = S_{\nu} D^2 / \kappa_d B_{\nu,T},
\label{eq:mdust}
\end{equation}

\noindent where $S_{\nu}$ is the measured flux density at frequency
$\nu$, $D$ is the distance to the galaxy of 675~kpc, $\kappa_d$ is the
grain mass absorption coefficient for which we assume a value of 1.2
m$^2$ kg$^{-1}$ at 160\mum\ (Li \& Draine 2001), and $B_{\nu,T}$ is
the Planck function. If the dust temperature is assumed to be
$\sim$~20K, this upper limit on the dust mass is $4.5 \times 10^{2}
\rm{M}_\odot$. This value is consistent with the upper limit of $4.1
\times 10^2 \rm{M}_\odot$ (and a temperature of $\sim$~20K) derived by
Temi et al.\ (2004) from ISOPHOT observations using their upper limit
on the flux density at 170\mum\ of 0.21~Jy. Using a gas-to-dust mass
ratio of 160, we infer an upper limit of the gas mass for the
``Total'' region of $7.2 \times 10^{4} \rm{M}_\odot$. This value is a
factor of ten more than the upper limit of $9.9 \times 10^3
\rm{M}_\odot$ derived by Sage, Welch \& Mitchell (1998). It is however
a factor of ten less than the estimated return of gas from evolved
stars to the ISM of $5.5 \times 10^5 \rm{M}_\odot$ calculated using
the prescription of Faber \& Gallagher (1976) for a time of $5 \times
10^8$~yr.

\subsection{NGC~185}

The best SED fits to the data for each region of NGC~185, which
produce the dust masses given in Table~8 and 9, are shown in
Figure~18. We measure a total dust mass for the ``Total'' region of
$1.9 \times 10^3 \rm{M}_\odot$. This estimate agrees with the value of
$1.5 \times 10^3 \rm{M}_\odot$ (at a temperature of 22~K) derived by
Temi et al.\ (2004) from ISO observations. Using a gas-to-dust mass
ratio of 160, we infer a gas mass for the ``Total'' region of $3.0
\times 10^5 \rm{M}_\odot$. This value is also in agreement, within a
factor of $\sim$3, with the value of $7.3 \times 10^5 \rm{M}_\odot$
derived by Sage, Welch \& Mitchell (1998) and the estimated return of
gas from evolved stars to the ISM of $8.4 \times 10^5 \rm{M}_\odot$
for an intermediate-age population $1 \times 10^9$~yr old.

Note that the best fit age for the ``Total'' region of
$450_{-330}^{+550}$ differs greatly from the other regions, which are
best fit assuming radiation fields produced by a very young
population. This is probably due to the fact that the ``Total'' region
contains emission from the entire galaxy and therefore more of an
older population component.

Note also that no temperatures are reported in Table~9. The dust
emission SED in our fitting procedure is the total emission from an
ensemble of grains with a size distribution and hence a temperature
distribution. Therefore, there is no single temperature for the
ensemble, even for those components in thermal equilibrium (large
graphitic and silicate grains). For the stochastically heated
components (PAH, small graphitic and silicate grains), speaking of a
temperature even for a single grain size within the ensemble makes
little sense. However, we can use a single-temperature fit to the 70
and 160\mum\ data for the ``Total'' region to obtain an estimate fo
the dust temperature, assuming a modified black body emissivity
function of the form:

\begin{equation}
F(\lambda) = \lambda^{-\beta} \frac{2 h c^2 \lambda^{-5}}{e^{h c / k \lambda T} - 1}.
\label{eq:mbb}
\end{equation}

\noindent Adopting an emissivity coefficient $\beta=2$ (valid beyond
70\mum; cf.\ Draine 2003), we estimate a temperature of $\sim$~17K.

\section{Spectral Properties of NGC~185's Dust Clouds}

\subsection{Spectral Line Measurements}

The remaining artifacts left in the IRS spectra, after extraction with
SPICE, were cleaned using the interactive analysis and processing tool
SMART\footnote{
http://ssc.spitzer.caltech.edu/archanaly/contributed/smart/} designed
for this purpose. The lines and features were then measured using
PAHFIT\footnote{http://tir.astro.utoledo.edu/jdsmith/pahfit.php}
(Smith et al.\ 2007) to analyze IRS low spectral resolution data in an
automated way. Despite the low resolution, PAHFIT tries to fit
important blended lines and features, for example [OIV] 25.890\mum\
and [Fe~II] 25.988\mum, and provides useful diagnostics in the
process, e.g.\ uncertainty on the fitted central wavelength. In those
cases where these uncertainties were relatively large, a cross check
was done manually using SMART. PAHFIT takes also into account the
extinction arisen either from a simple fully-mixed or screen dust
component, dominated by the silicate absorption bands at 9.7 and 18
microns.  For our measurements we have chosen the most simple ``screen
dust extinction'' approximation to implement a reasonable correction
to the line fluxes. The measurements resulting from the PAHFIT are
given in Table~10 and 11.

The wavelength coverage of the IRS spectra is nominally 5 to 40\mum,
although at longer wavelengths, in particular for relatively faint
emission, some artifacts like fringing and noisy pixels, can affect
the spectrum and therefore a reliable line identification is an
issue. The useful wavelength (IRS Data Handbook Version 3.1, 2007)
range is 5.2 to 37\mum. This is relevant because some expected lines
from photodissociation regions (PDR), like H$_2$ S(7) 5.51\mum\ and
[Fe~II] 35.35 and 35.78\mum\ fall near these edges and are not
necessarily automatically fitted by PAHFIT. The fits are done at the
native spectral resolution and the plotted spectra has been smoothed
out using a boxcar filter with 3 pixels width, just slightly wider
than a resolution element. The final spectra, along with identified
lines and features, are shown in Figure~19. We notice that for a few
emission lines the rest wavelength found by PAHFIT (Table~10) is off
by more than 0.5\mum. This typically takes place in regions of the
spectrum where the PAH emission is strong or near the edge of the
spectrum (e.g.\ for the [Si~II] 34.8\mum\ line).

\subsection{Analysis of the Spectra}

NGC~185 has been classified as a low luminosity Seyfert type 2 galaxy,
based on its optical and radio properties (Panessa et al.\ 2006; Nagar
et al.\ 2005), although the presence of a black hole as a possible
driver of this activity has not been confirmed in X-ray observations
(Brandt et al.\ 1997).

The analysis of the NGC~185 IRS spectra is relatively straightforward,
and can be placed in a more global context, thanks to recent works on
normal galaxies by the SINGs Legacy team (Draine et al.\ 2007; Roussel
et al.\ 2007), or Seyfert galaxies (Buchanan et al.\ 2006), based on
Spitzer data and IRS measurements. Although there are some difference
in the IRS spectra between the three different observed positions,
some trends are very similar: strong PAH emission, deep silicate
absorption bands at $\sim$~9.7 and 18\mum, the presence of relative
low excitation fine structure atomic emission lines, (except perhaps
by the presence of [O~IV] 25.9\mum) as well as that of the 0-0 S(0)
and S(1) H$_2$ rotational lines.  These features are clear indicators
of the presence of dust and ultraviolet ionization from star formation
activity. The spectra overall resemble that of a photodissociation
region, and in this sense they look very much like that of some of
object studied by Buchanan et al.\ (2006) in their sample of
PAH-dominated Seyfert galaxies, e.g.\ NGC~3079. Note that NGC~3079,
with its very red spectrum and strong PAH emission, shows a ``bubble''
at its center and that this bubble was probably created by winds
released during a burst of star formation (Cecil et al.\ 2001).

PAHs provide an interesting diagnostic to classify galaxies emitting
in the mid-infrared when they are compared to the silicate 9.78\mum\
absorption feature that enables one to distinguish between AGNs and
starburst-dominated systems (Spoon et al.\ 2007). As pointed out
before, the NGC~185 spectra show deep silicate absorption at 9.7 and
18\mum\ as well as strong 6.2\mum\ PAH emission. For our ``North'',
``Center'' and ``South'' target positions, we measured the
equivalent width (EW) of the 6.2\mum\ PAH emission feature ($0.7 \pm
0.2$, $0.4 \pm 0.1$ and $0.8 \pm 0.2$, respectively) as well as the
strength of the 9.7\mum\ silicate feature ($-1.0 \pm 0.2$, $-1.3 \pm
0.3$ and $-2.6 \pm 0.3$, respectively) and plotted the two quantities
in the diagnostic diagram of Spoon et al.\ (2007; Figure~1). The flux
in the 6.2\mum\ PAH emission band is measured by integrating the flux
above a spline interpolated local continuum from 5.95 to 6.55\mum.
The EW of the PAH feature is then obtained by dividing the integrated
PAH flux by the interpolated continuum flux density below the peak
($\sim$~6.22\mum) of the PAH feature. The apparent strength of the
9.7\mum\ silicate feature is inferred by adopting a local mid-infrared
continuum and evaluating the ratio of observed flux density
($f_{obs}$) to continuum flux density ($f_{cont}$) at 9.7\mum, as
defined in Spoon et al.\ (2007):

\begin{equation}
S_{sil} = ln \frac{f_{obs} (9.7\mu \rm{m})}{f_{cont} (9.7\mu \rm{m})}.
\label{eq:ssil}
\end{equation}

\noindent The error bars were calculated from different measurements
for slightly different fit to the continuum baseline. We did not apply
any correction for possible water ice absorption features.

Based on these numbers, our spectra fall into the class 1C-2C
(PAH-dominated spectra), 2B-2C (weaker PAH features than class 2C) and
2C. All spectra occupy part of the diagram that is populated mainly by
starbust galaxies. The Seyfert galaxies and QSOs are generally
classified as class 1A which are characterized by a nearly featureless
hot dust continuum with a very weak silicate absorption feature at
9.7\mum.  Therefore, based on its infrared properties, NGC~185 appears
to be currently forming stars (with lifetimes of
$\lesssim$~20~Myr). If this interpretation is correct, this relatively
young stellar population has remained undetected until now, possibly
due to the large extinction due to the large amount of dust in this
galaxy.

If the PAH emission is indeed being excited by far-ultraviolet (FUV)
radiation from young stars, then we can estimate the current star
formation rate based on a relationship derived from recent surveys of
galactic and extragalactic objects (Wu et al.\ 2005; Peeters
2004). Using the PAH emission at 7.7\mum\ the star formation rate
(SFR) can be estimated (Wu et al.\ 2005) to be,

\begin{equation}
\rm{SFR}_{8\mu \rm{m} (dust)} = \frac{\nu L_\nu
[8\mu \rm{m} (dust)]}{1.57 \times 10^9 \rm{L}_\odot},
\label{eq:sfrpah}
\end{equation}

\noindent where $\nu L_\nu$ is in L$_\odot$ and the SFR is given in
M$_\odot$/yr. Using the contribution of the three IRS target positions
to the 7.7\mum\ PAH feature computed from the ``Total'' 8\mum\ flux
density value in Table~6 (92~mJy), we estimate a current SFR~$\sim 5
\times 10^{-11} \rm{M}_\odot$/yr. Clearly this is a lower limit but
not too far from what has been inferred before by Butler \&
Martinez-Delgado (2005), i.e.\ a mean star formation density rate of
at least $2.6 \times 10^{-9} \rm{M}_\odot$ yr$^{-1}$ pc$^{-2}$ within
the central 2\arcmin\ over $10^9$ yr.

The H$_2$ 0-0 S(0), 0-0 S(1), 0-0 S(2), and 0-0 S(3) pure rotational
line emission arising from the PDR has been estimated by Kaufman,
Wolfire \& Hollenbach (2006) and depends on the PDR temperature, the
hydrogen nucleus density $n$ (in units of cm$^{-3}$) and the FUV
radiation field strength $G_0$ (in units of $1.6 \times
10^{-3}$~erg~cm$^{-2}$~s$^{-1}$, the local ISM strength of the diffuse
radiation field; Habing 1968). We can compare our line ratios, for
instance the ratio of the H$_2$ S(1) 17.03 and S(0) 28.22\mum\ lines,
to the predicted line ratios of Kaufman, Wolfire \& Hollenbach
(2006). For the three observed positions (``North'', ``Center'' and
``South'', respectively) our calculated ratios of S(1)$/$S(0) are 2.1,
3.6 and 2.4, corresponding to a FUV radiation field in the range $2.0
\le$~log~(G$_0$)~$\le 5.0$ and a hydrogen nucleus density of $2.6
\le$~log~(n)~$\le 5$. These values are therefore consistent with
having the molecular gas inside the dust cloud being impinged by the
FUV radiation field of a relatively young stellar population in
NGC~185.

In the outer layers of PDRs (A$_{V} \le$ 1), one expects emission from
the [Si~II] 34.8\mum\ and [Fe~II] 26\mum\ transitions (Kaufman,
Wolfire \& Hollenbach 2006). The ``North'' and ``South'' positions
show clear signs of the [Si~II] 34.8\mum\ line, and very likely
[Fe~II]. The presence of [O~IV] 25.89\mum\ in Seyfert galaxies using
low resolution IRS spectroscopy has been recently discussed and
analyzed on a relatively large sample of objects ($>$ 50; Mel\'endez
et al.\ 2008). The point has been made that the possible blending of
the [Fe~II] emission line at 26\mum\ is an issue, in particular for
the less energetic objects. Because of this we have left PAHFIT to
carry out the deblending of these two lines automatically and have
kept both entries in our measurements.

Archival observations (from January 2004) in the ultraviolet (UV) by
the XMM-OM instrument strongly suggest recent star formation activity
in NGC~185 (see Figure~20). The XMM-OM camera was used with the UVW1
filter ($2330 - 2950$\AA) and therefore overlaps in wavelength
coverage with the GALEX NUV channel ($1800 -2750$\AA). This means that
the UVW1 image can be used equally well as a reliable tracer of star
formation. The XMM-OM camera has a 17\arcmin\ field-of-view with an
angular resolution in the UVW1 band comparable to that of the IRAC
bands (FWHM $\sim$ 2\arcsec) and a sensitivity that matches within a
factor of two that of GALEX (Morrissey et al.\ 2005; Mason et al.\
2001). Figure~20 shows a comparison of the XMM-OM UVW1 observation
(left) with those of {\it Spitzer} at 8\mum\ (center) and 24\mum\
(right) covering the central two arcminutes of NGC~185. The grayscale
is such that bright regions are dark and extincted regions white. The
brightest sources in the UVW1 image have been subtracted to emphasize
the diffuse UV emission although some subtraction residuals are still
present. The ``North'' dust cloud, located $\sim 15\arcsec$\ from the
center, is quite distinguishable in the UV, as well as a fainter
bridge that connects it with the ``South'' dust cloud. The white
contours in the {\it Spitzer} images correspond to that of the UV
emission which is essentially found at the center of the galaxy and
surrounded by the dust clouds. Our spectroscopic observations have
been positioned over the dust cloud purposely, therefore the fact that
strong PAH emission is detected at these positions suggests that {\it
locally}, i.e.\ not too far from the clouds, there is a source of UV
photons interacting with them. Nevertheless, it is certainly possible
that UV light from other sources (e.g.\ planetary nebulae) could also
contribute to the excitation. If this is the case, this would affect
the star formation rate computed above based on the 8\mum\ emission.

In Figure~21, we compare the average of the three IRS spectra with the
measured and best fit SED for ``NGC185Total''. The average spectrum
was normalized to match the SED model in the $20-30$\mum\ wavelength
region, where the stellar photometric contribution is the lowest. The
figure also includes both the ISOCAM and ISOPHOT measurements of
Xilouris et al.\ (2004) for comparison. The ISO data points are for
measurements done at 4.5, 6.7, 15\mum\ with ISOCAM, and at 60, 90, 170
and 200\mum\ with ISOPHOT. There is remarkably good agreement between
the best fit SED and the ISO data points even though the actual model
fits were done using only the 2MASS and Spitzer data. The PAH features
used in the model agree remarkably well with the data. At shorter
wavelengths, the ``NGC185Total'' emission comes mainly from stars and
therefore our average spectrum of the dust clouds falls below the
model SED.

\section{Summary}

New IRAC, MIPS and IRS observations of NGC 147 and NGC~185 give a
better assessment of the dust content and properties in these
prototypical local dwarf galaxies. Spitzer's high sensitivity and
spatial resolution enable us for the first time to look directly into
the detailed spatial structure and properties of the dust in these
systems. The images of NGC~185 at 8 and 24 micron display a mixed
morphology characterized by a shell-like diffuse emission region
surrounding a central concentration of more intense infrared
emission. The lower resolution images at longer wavelengths show the
same spatial distribution within the central 50\arcsec\ but beyond
this radius, the 160\mum\ emission is more extended than that at 24
and 70\mum. On the other hand, the dwarf galaxy NGC 147 located only a
small distance away from NGC~185 shows no significant infrared
emission beyond 24 micron and its diffuse infrared emission is mainly
stellar in origin.

For NGC~147, we obtain an upper limit for the dust mass of $4.5 \times
10^{2} \rm{M}_\odot$, a value consistent with the previous upper limit
derived using ISO observations of this galaxy. For NGC~185, the
derived dust mass based on the best fit to the ``Total'' SED is $1.9
\times 10^3 \rm{M}_\odot$, implying a gas mass of $3.0 \times 10^5
\rm{M}_\odot$ (assuming a standard gas-to-dust mass ratio of
160). These values are in agreement with those previously estimated
from infrared as well as CO and HI observations. The gas estimate is
also consistent with the predicted mass return from dying stars, based
on the last burst of star formation, $1 \times 10^9$~yr ago. Based on
simply the 70 to 160\mum\ flux density ratio, the estimated
temperature for the dust is $\sim$~17K. The fact that NGC~147
resembles more a ``typical'' dust and gas-free elliptical galaxy than
NGC~185 (and NGC~205, for that matter, see Marleau et al.\ 2006)
remains puzzling in the context of the possible binary system
scenario.

In the case of NGC~185, we also presented full $5-38$\mum\
low-resolution (R$\sim$100) spectra of the main emission regions. The
IRS spectra of NGC~185 show strong PAH emission, deep silicate
absorption features and H$_2$ pure rotational line ratios consistent
with having the dust and molecular gas inside the dust cloud being
impinged by the far-ultraviolet radiation field of a relatively young
stellar population. Although the current rate of star formation is
quite low ($\sim 10^{-10} \rm{M}_\odot$/yr), this suggests that the
star formation history of NGC~185 is complex, perhaps as much as that
of the more active NGC~205 (Monaco et al.\ 2009).

\acknowledgements

This work is based on observations made with the Spitzer Space
Telescope, which is operated by the Jet Propulsion Laboratory (JPL),
California Institute of Technology, under NASA contract 1407. Support
for this work was provided by NASA contract NAS7-03001. We thank the
referee for her/his careful reading of our manuscript; her/his
comments have improved our presentation and discussion.

\newpage

\begin{figure}[h!]
\centerline{\hbox{
\includegraphics[width=270pt,height=480pt,angle=-90]{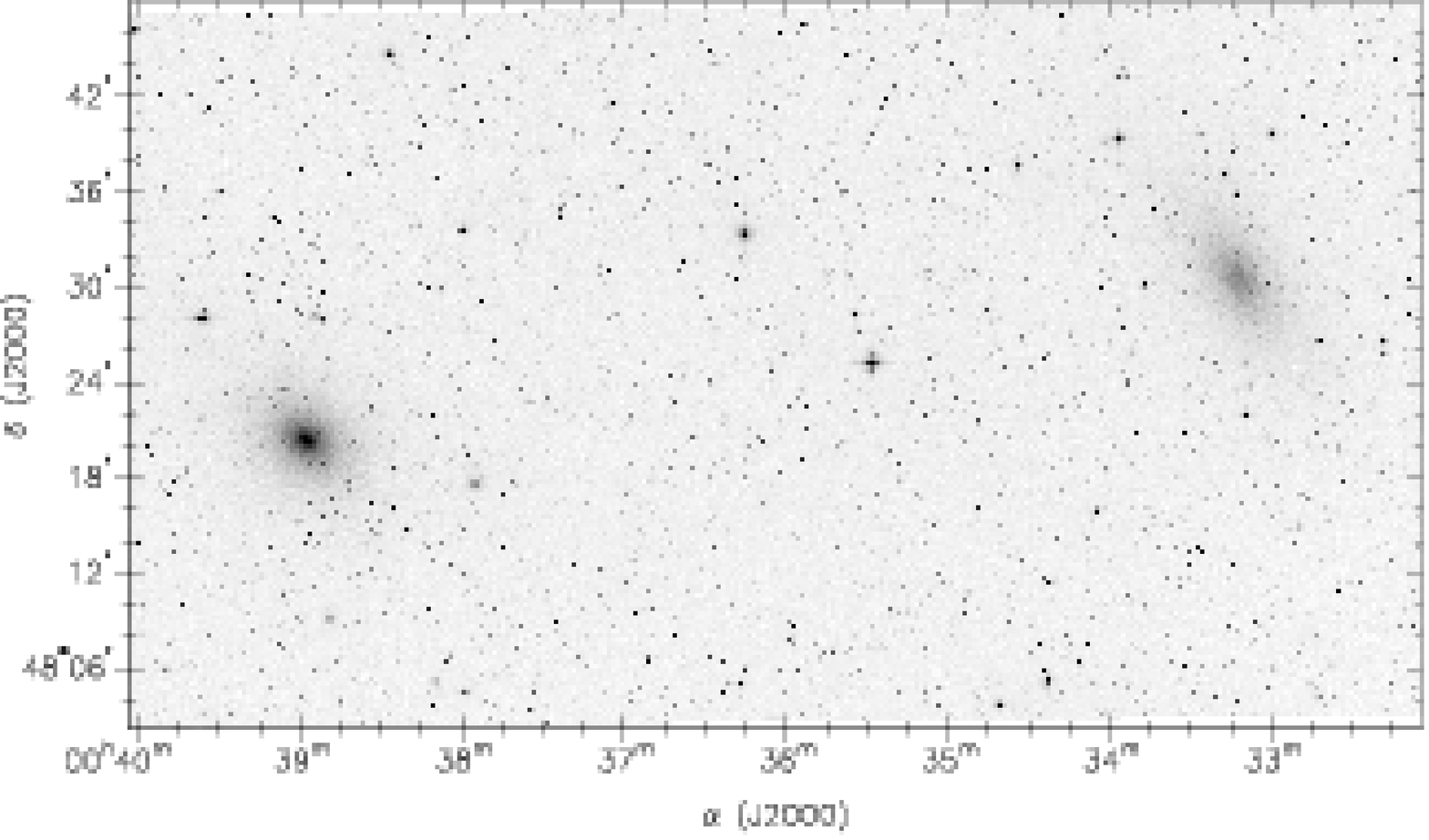}
}}
\caption{\label{fig:1} Digitized Sky Survey (DSS-2-red) image showing
the location on the sky of the two satellite galaxies NGC~185 ({\it
left}) and NGC~147 ({\it right}), with an angular separation of only
roughly one degree (or $\sim$~11.2~kpc, assuming an average distance
to these galaxies of $\sim$~645~Mpc).}
\end{figure}

\newpage
\begin{figure}[h!]
\centerline{\hbox{
\includegraphics[width=480pt,height=480pt,angle=-90]{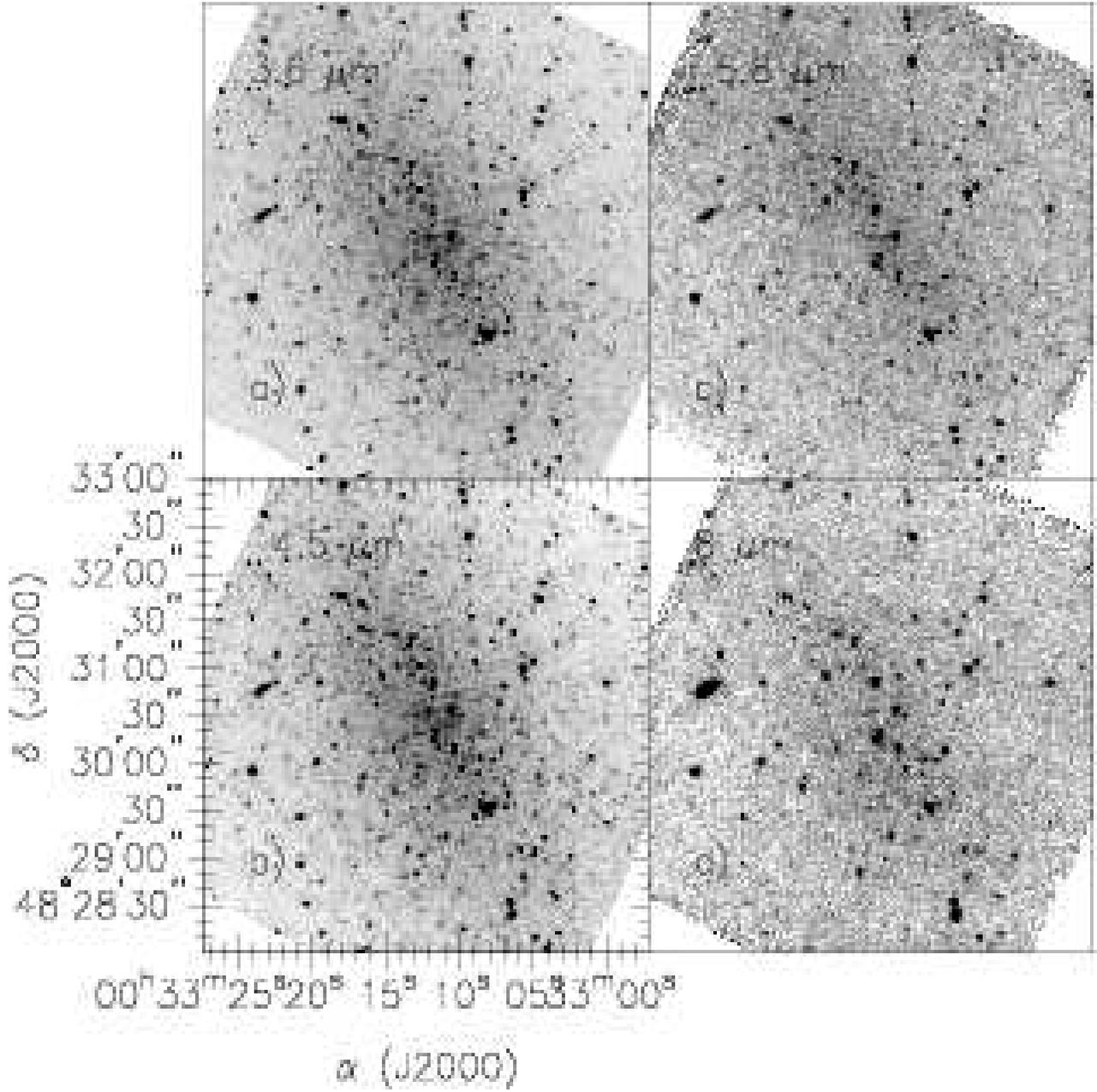}
}}
\caption{\label{fig:2} IRAC 3.6, 4.5, 5.8 \& 8\mum\ images of
NGC~147. The field-of-view (FOV) of each image is 5\arcmin\ $\times$
5\arcmin\ with N up and E to the left.}
\end{figure}

\newpage
\begin{figure}[h!]
\centerline{\hbox{
\includegraphics[width=480pt,height=480pt,angle=-90]{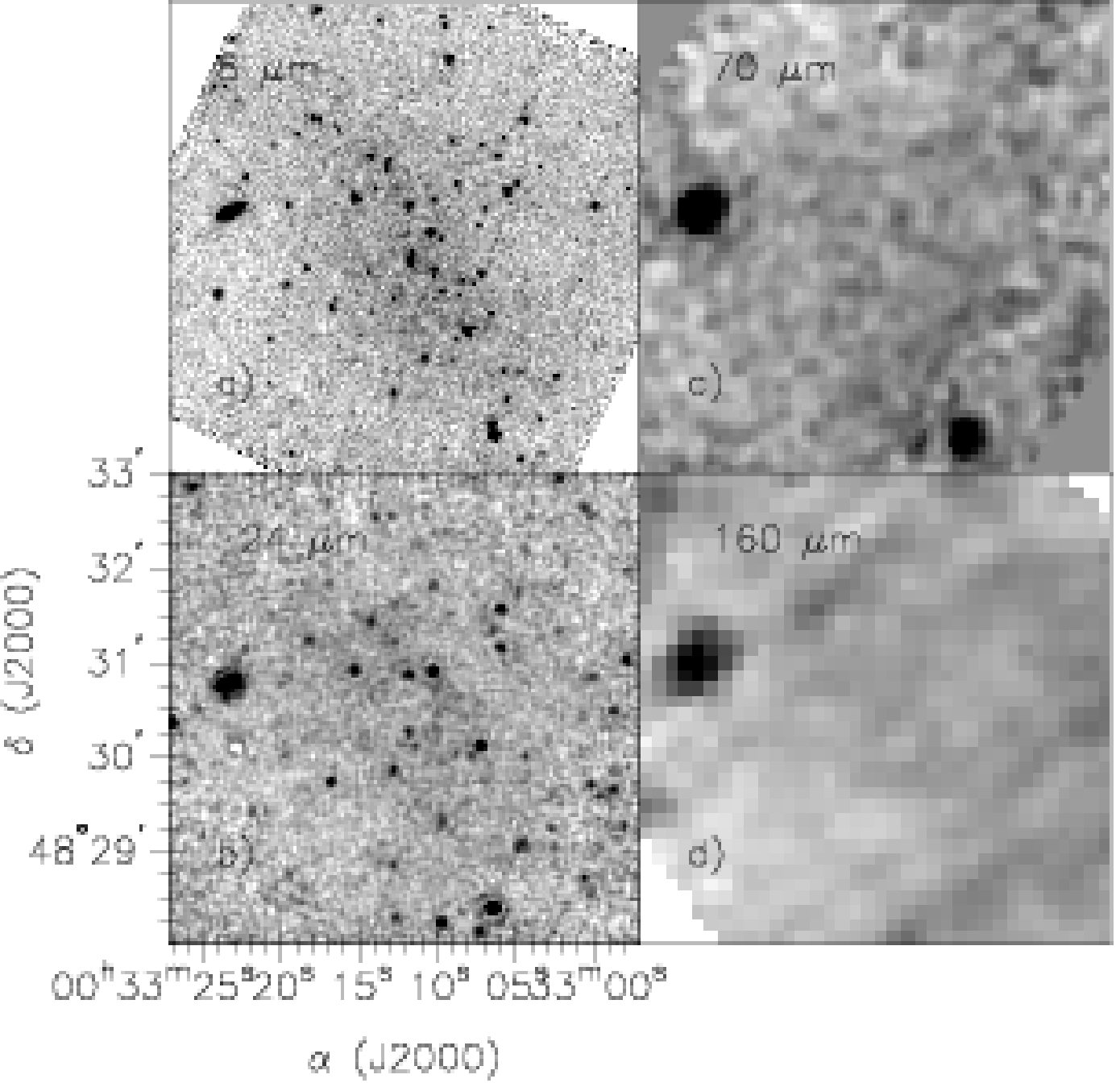}
}}
\caption{\label{fig:3} IRAC 8\mum\, MIPS 24, 70 \& 160\mum\ images of
NGC~147. The FOV of each image is 5\arcmin\ $\times$ 5\arcmin\ with N
up and E to the left.}
\end{figure}

\newpage
\begin{figure}[h!]
\centerline{\hbox{
\includegraphics[width=480pt,height=480pt,angle=-90]{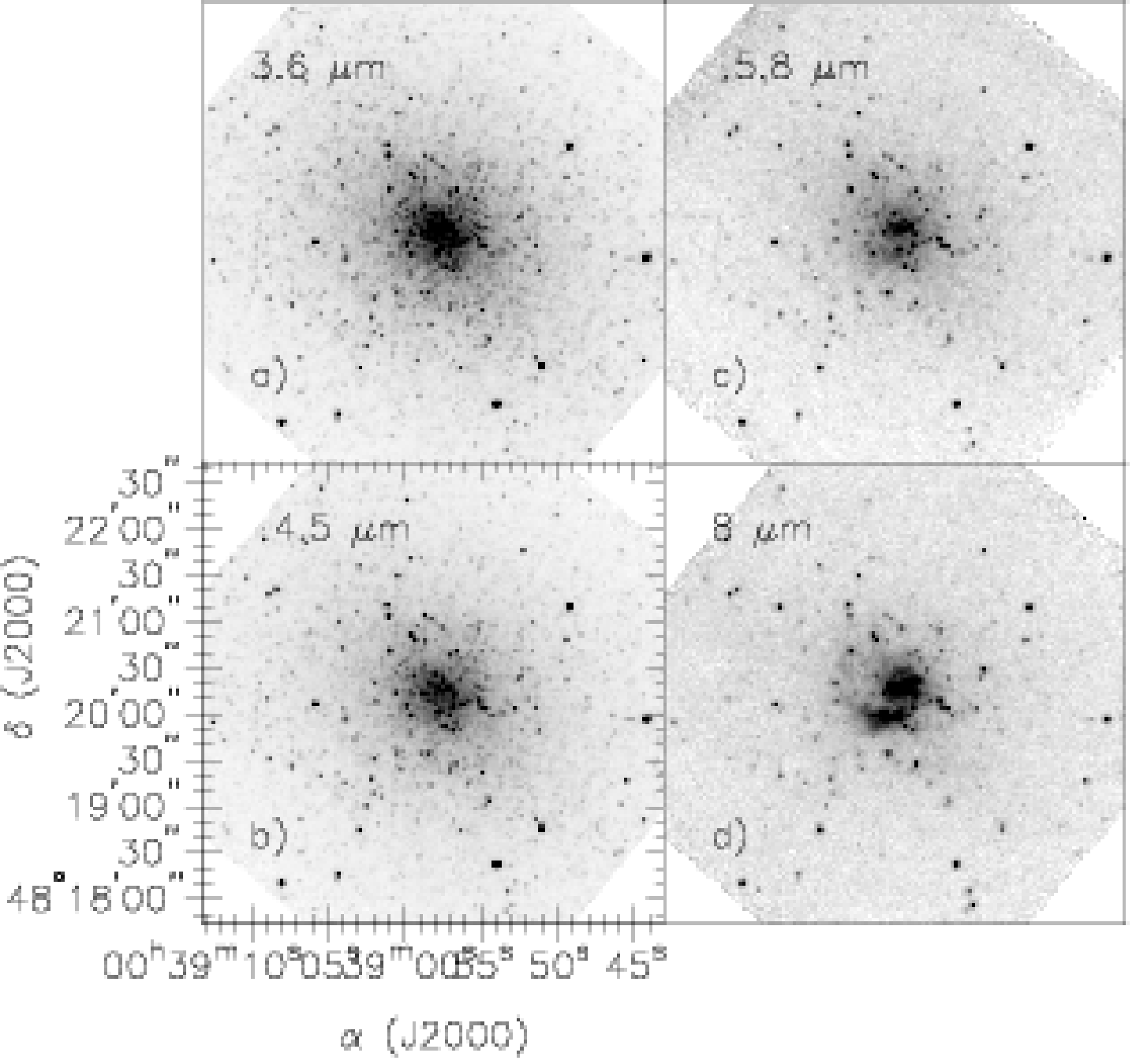}
}}
\caption{\label{fig:4} IRAC 3.6, 4.5, 5.8 \& 8\mum\ images of
NGC~185. The FOV of each image is 5\arcmin\ $\times$ 5\arcmin\ with N
up and E to the left.}
\end{figure}

\newpage
\begin{figure}[h!]
\centerline{\hbox{
\includegraphics[width=480pt,height=480pt,angle=-90]{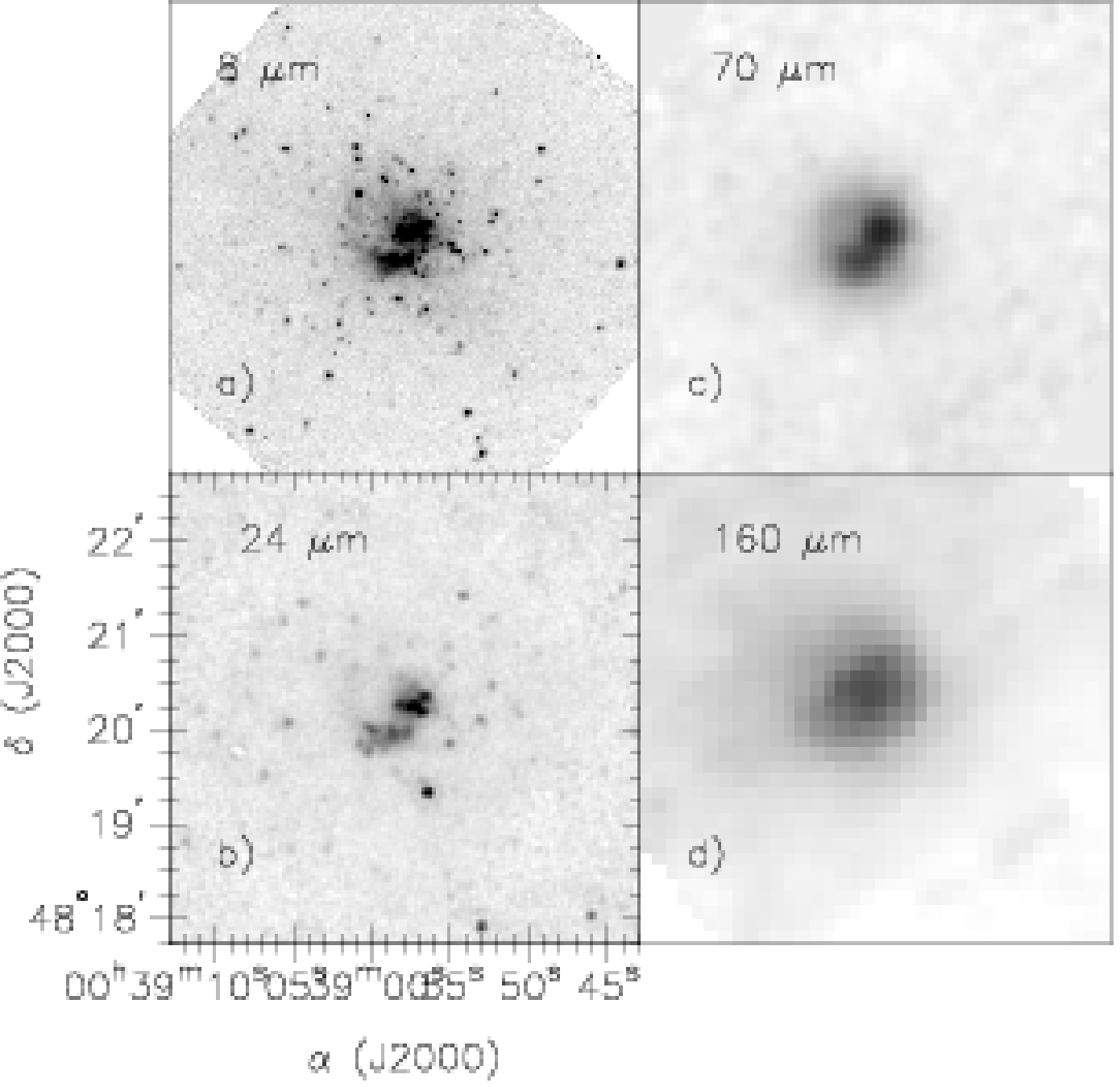}
}}
\caption{\label{fig:5} IRAC 8\mum, MIPS 24, 70 \& 160\mum\ images of
NGC~185. The FOV of each image is 5\arcmin\ $\times$ 5\arcmin\ with N
up and E to the left.}
\end{figure}

\newpage
\begin{figure}[h!]
\centerline{\hbox{
\includegraphics[width=400pt,height=400pt,angle=0]{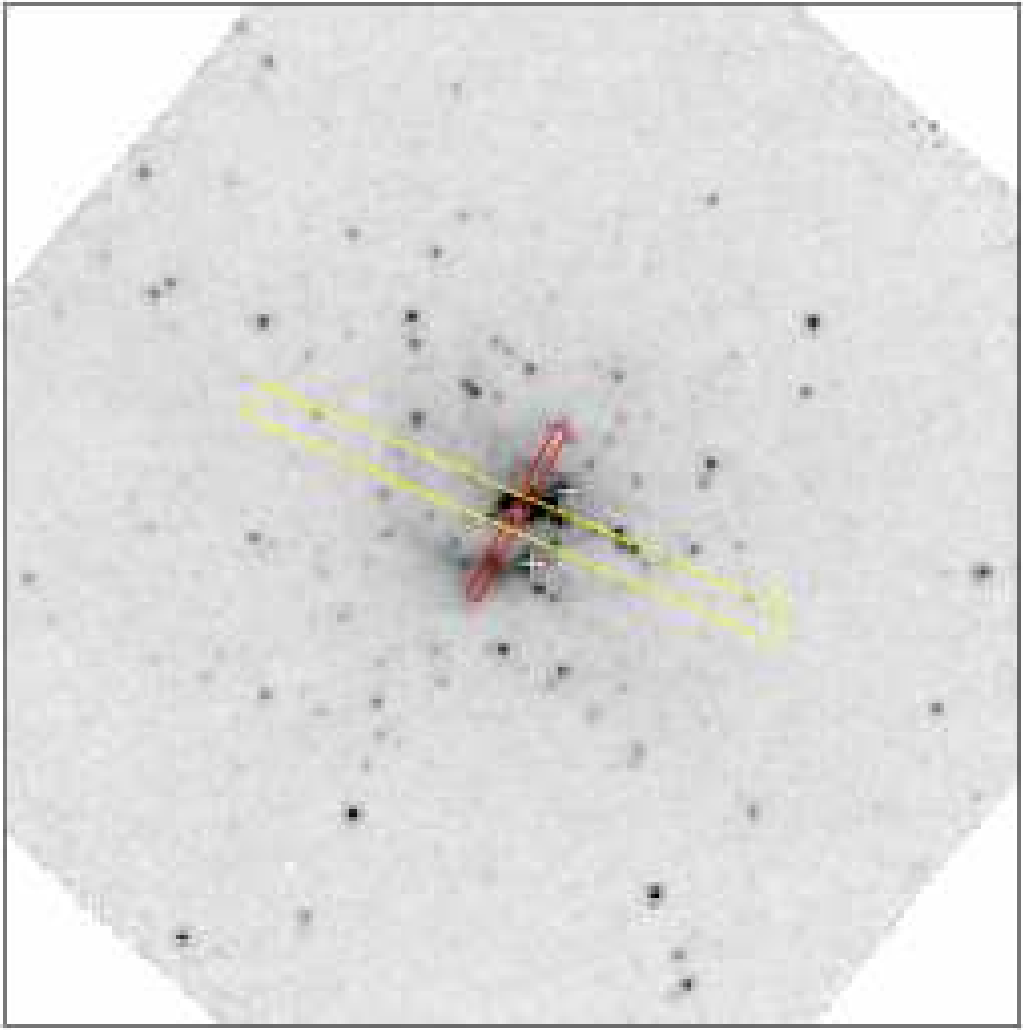}
}}
\caption{\label{fig:6} The three IRS target positions ({\em white
crosses}), as given in Table~1, are shown overlayed on the IRAC
8\mum\ image of NGC~185.  The image is 5\arcmin\ on a side, with N up
and E to the left.  The IRS Short-Low (SL, {\em red}) and Long-Low
(LL, {\em yellow}) slits are also displayed on the same image for one
of the target position (NGC185IRSCenter).}
\end{figure}

\newpage
\begin{figure}[h!]
\centerline{\hbox{
\includegraphics[width=120pt,height=120pt,angle=0]{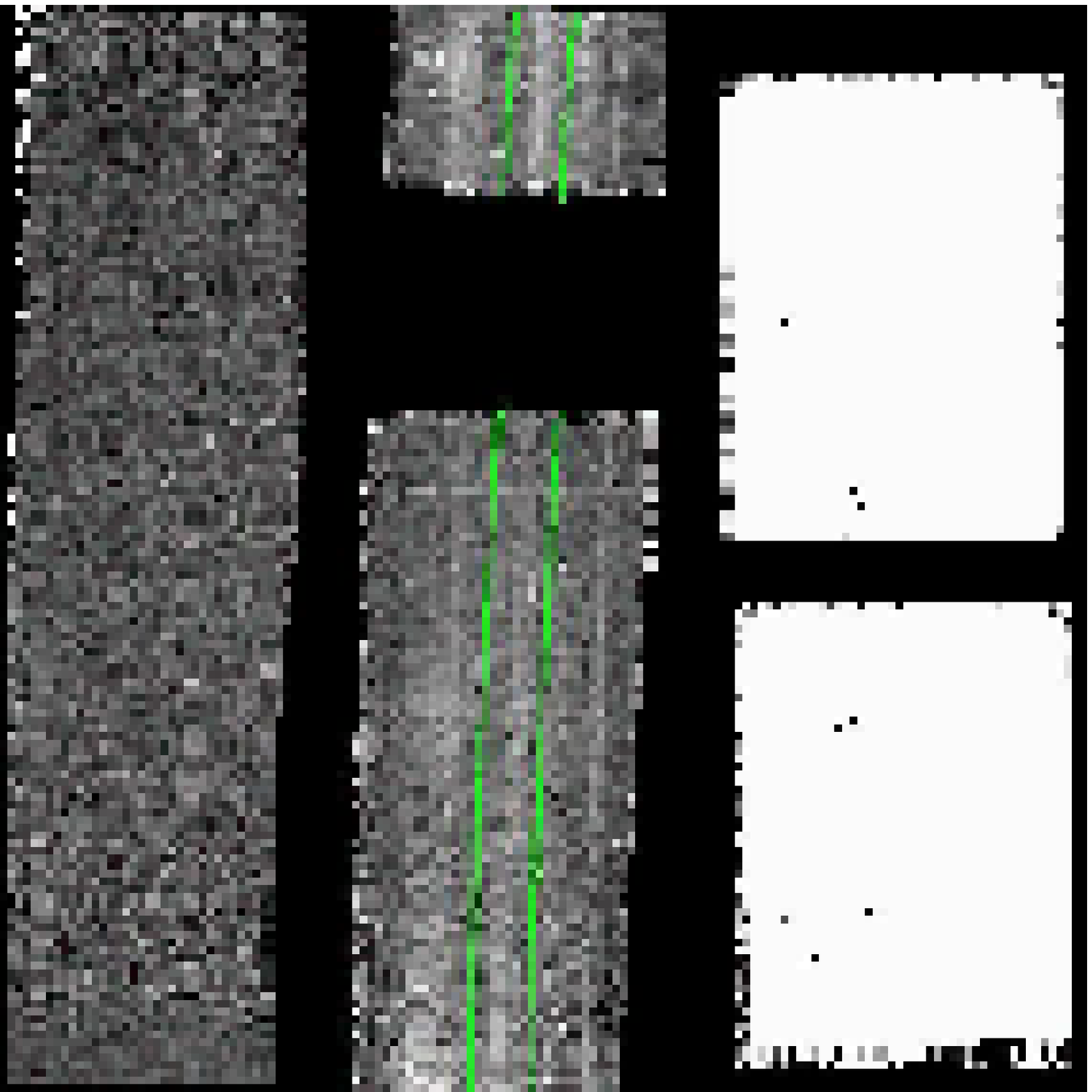}
\includegraphics[width=120pt,height=120pt,angle=0]{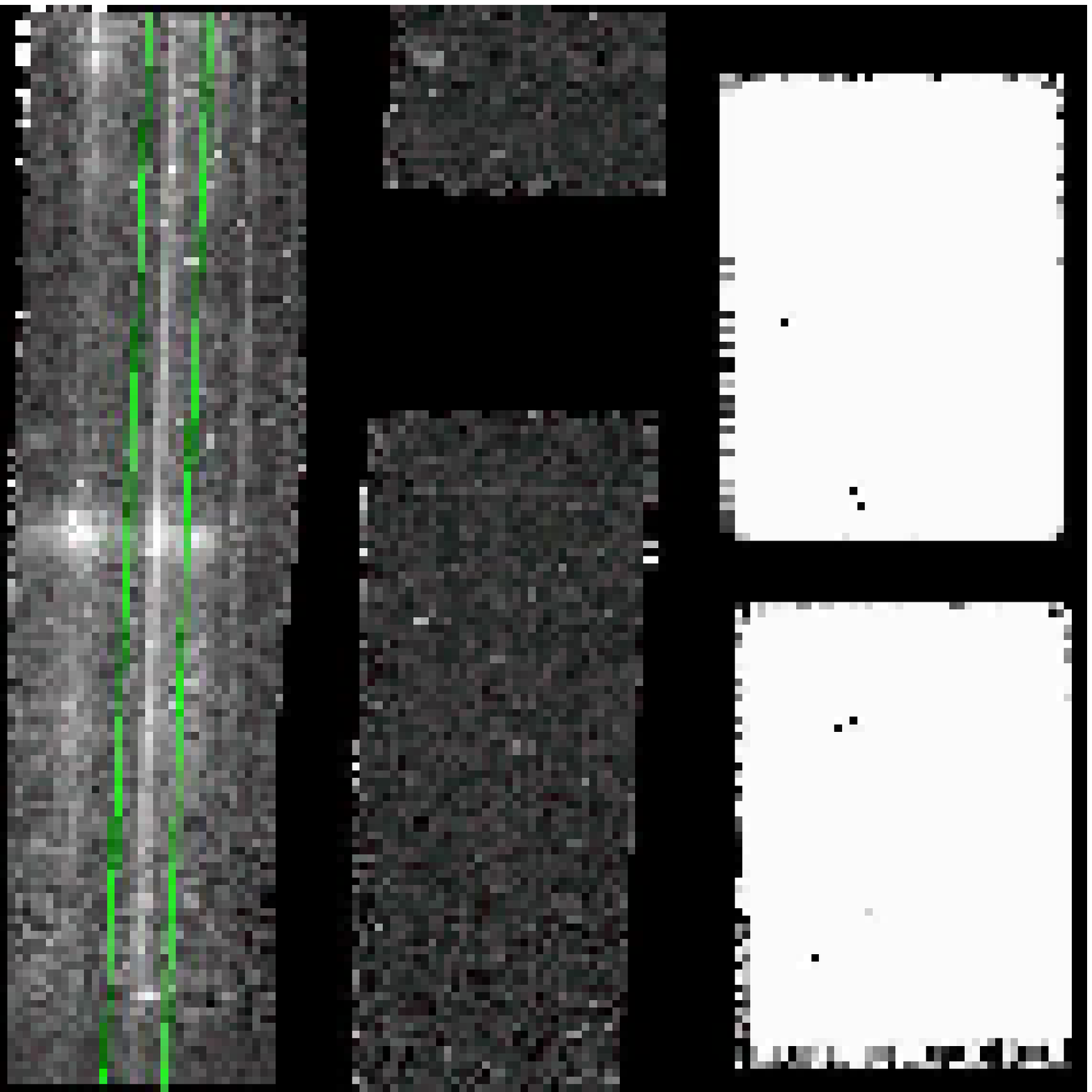}
\includegraphics[width=120pt,height=120pt,angle=0]{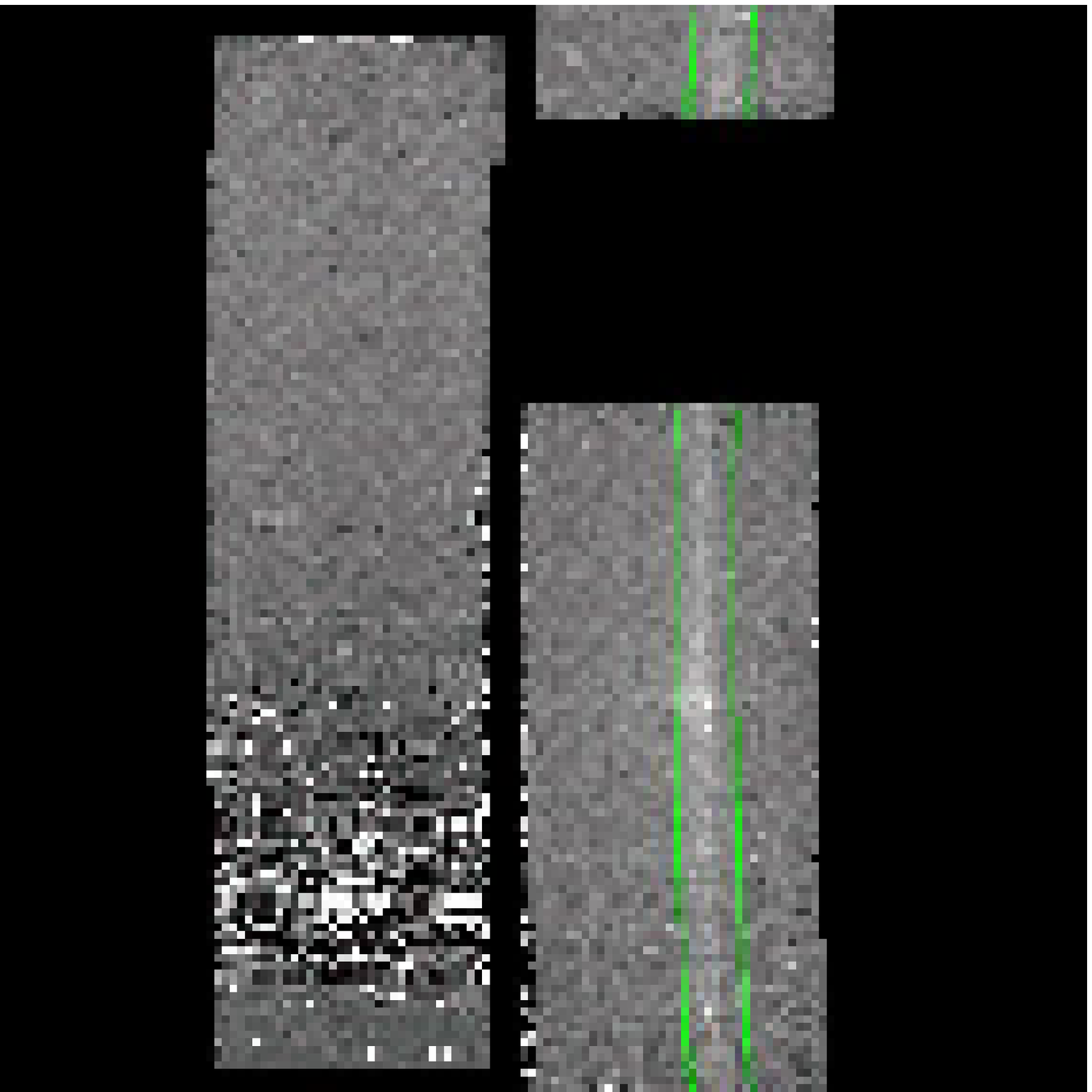}
\includegraphics[width=120pt,height=120pt,angle=0]{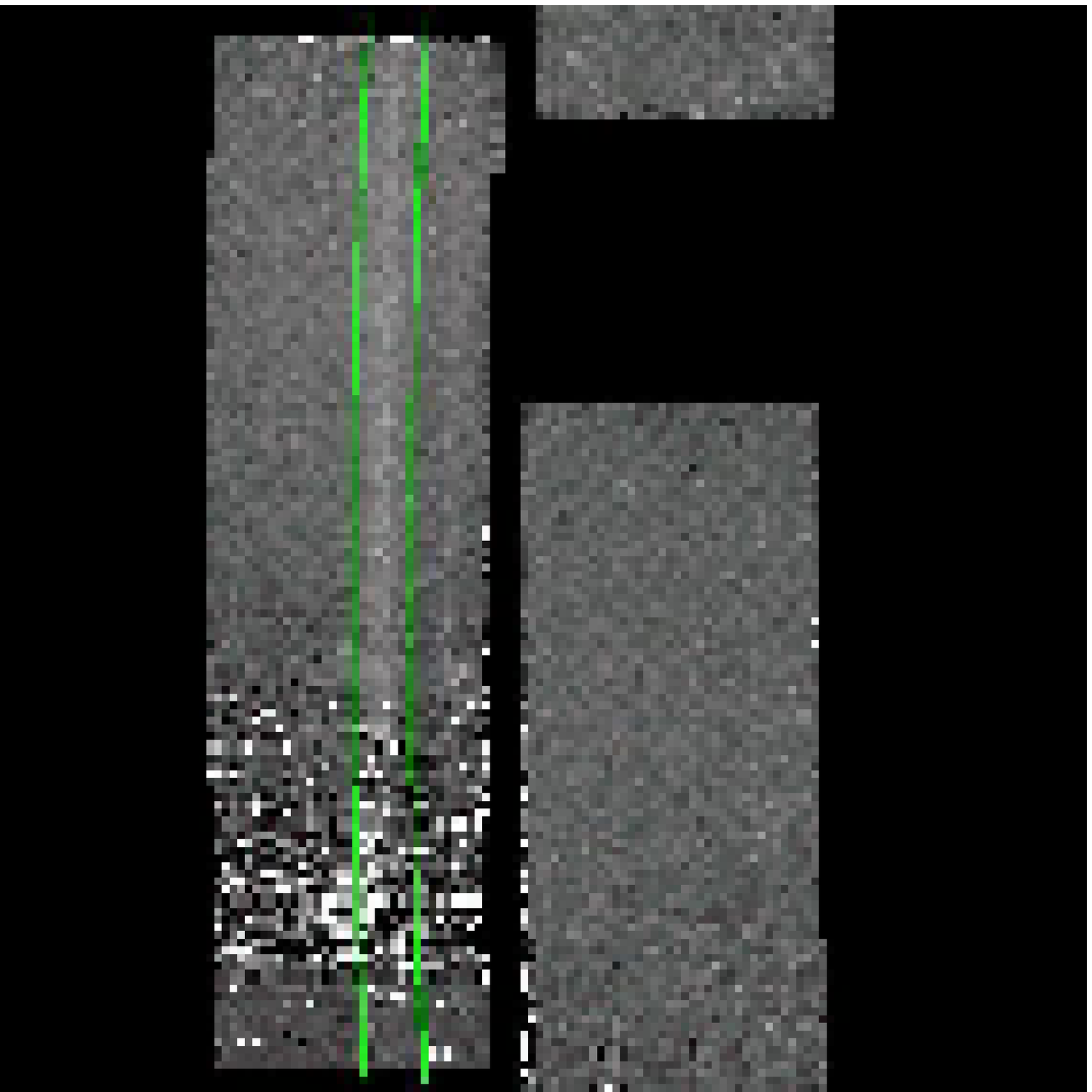}
}}
\centerline{\hbox{
\includegraphics[width=120pt,height=120pt,angle=0]{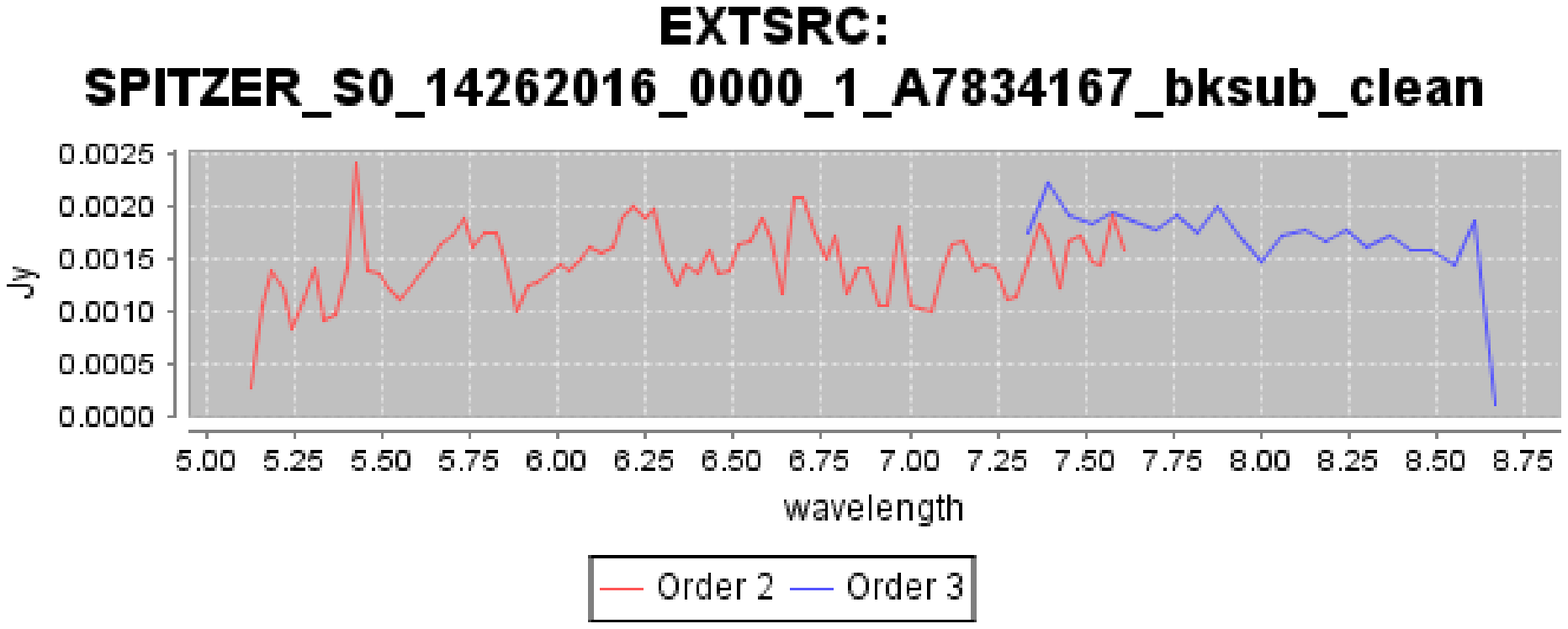}
\includegraphics[width=120pt,height=120pt,angle=0]{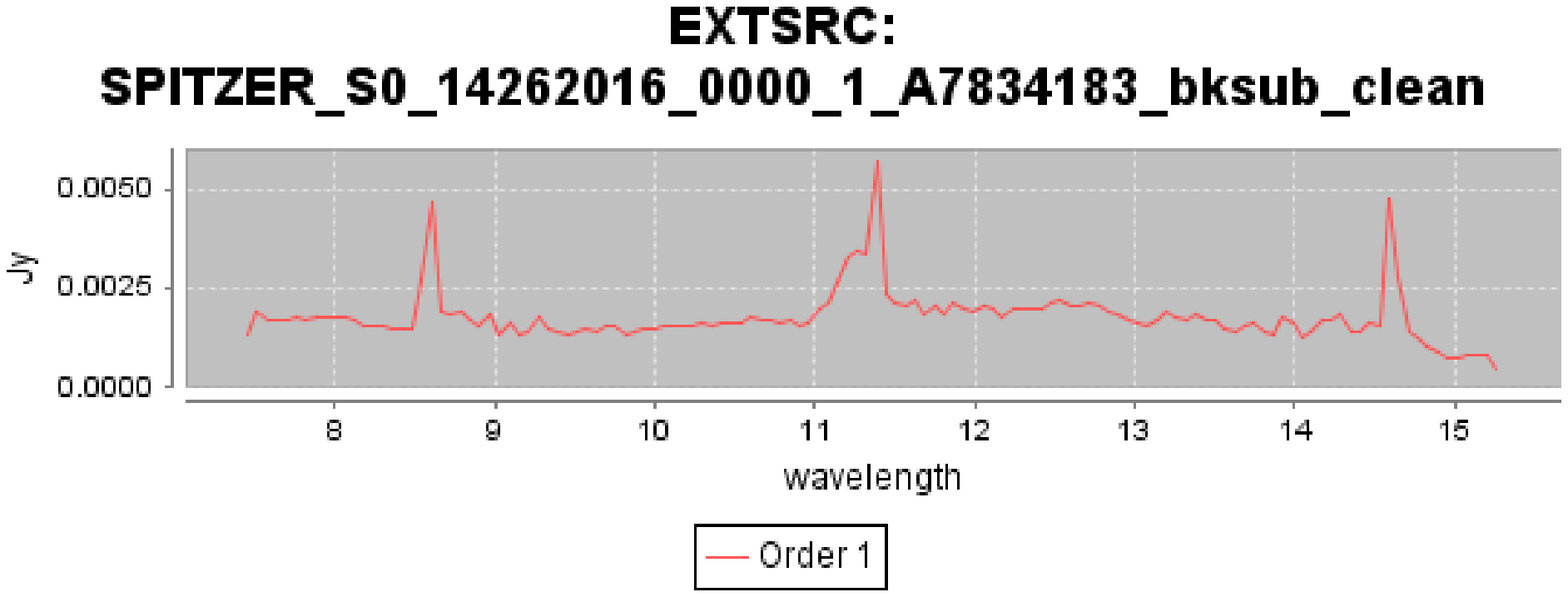}
\includegraphics[width=120pt,height=120pt,angle=0]{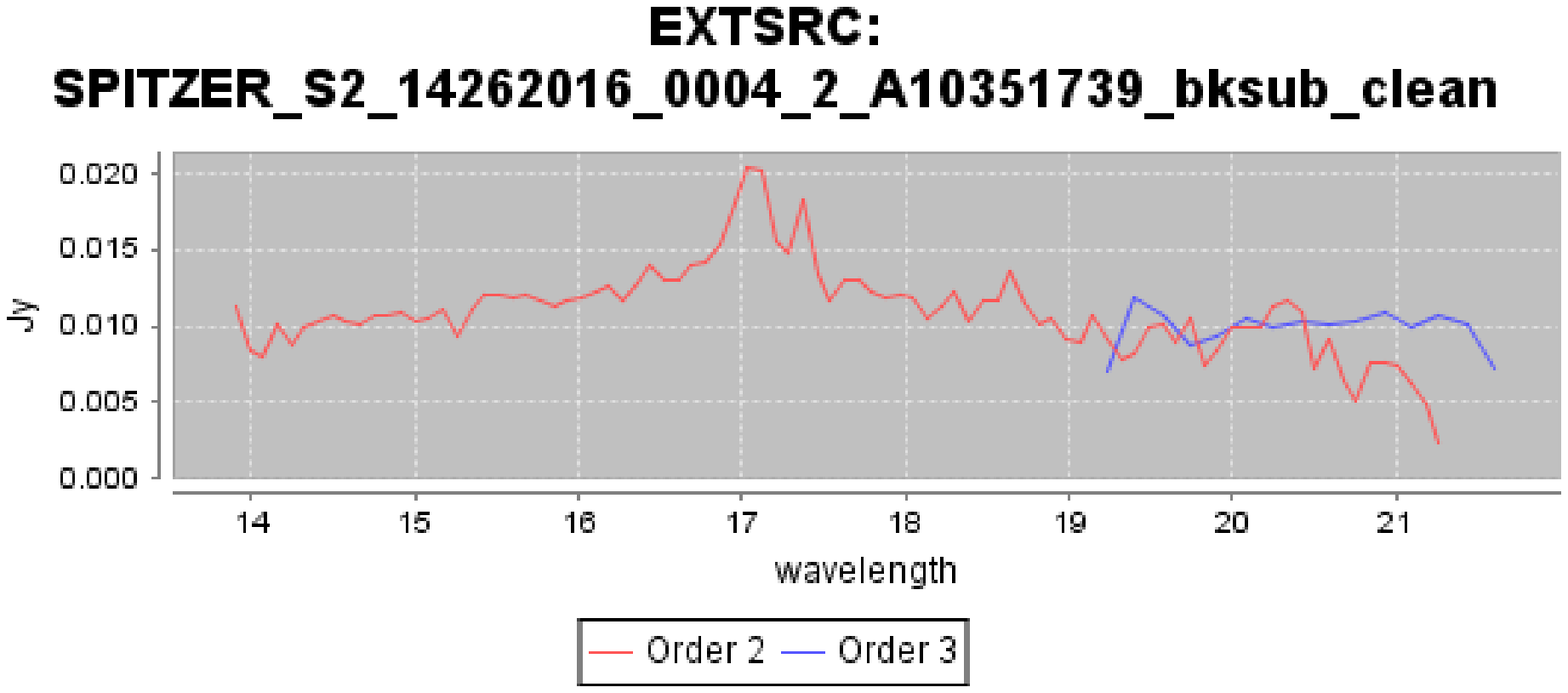}
\includegraphics[width=120pt,height=120pt,angle=0]{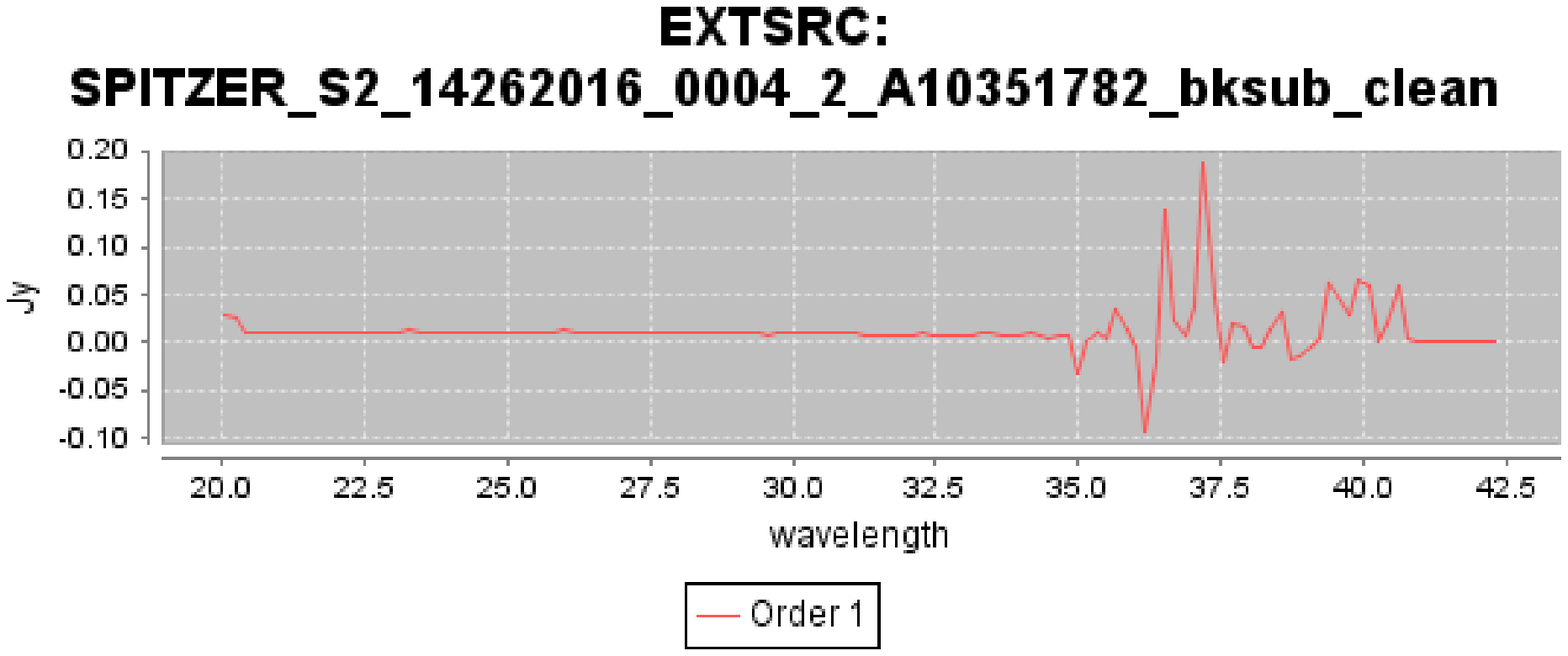}
}}
\caption{\label{fig:7} {\em Top}: Example of the spectral images taken
at the ``NGC185IRSNorth'' (see Table~1) target position for, {\em from
left to right}: the SL 2nd order, SL 1st order, LL 2nd order and LL
1st order. The {\em green lines} shows the region of extraction. {\em
Bottom}: Extracted spectra corresponding to the spectral images shown
at the top.}
\end{figure}

\newpage
\begin{figure}[h!]
\centerline{\hbox{
\includegraphics[width=400pt,height=400pt,angle=0]{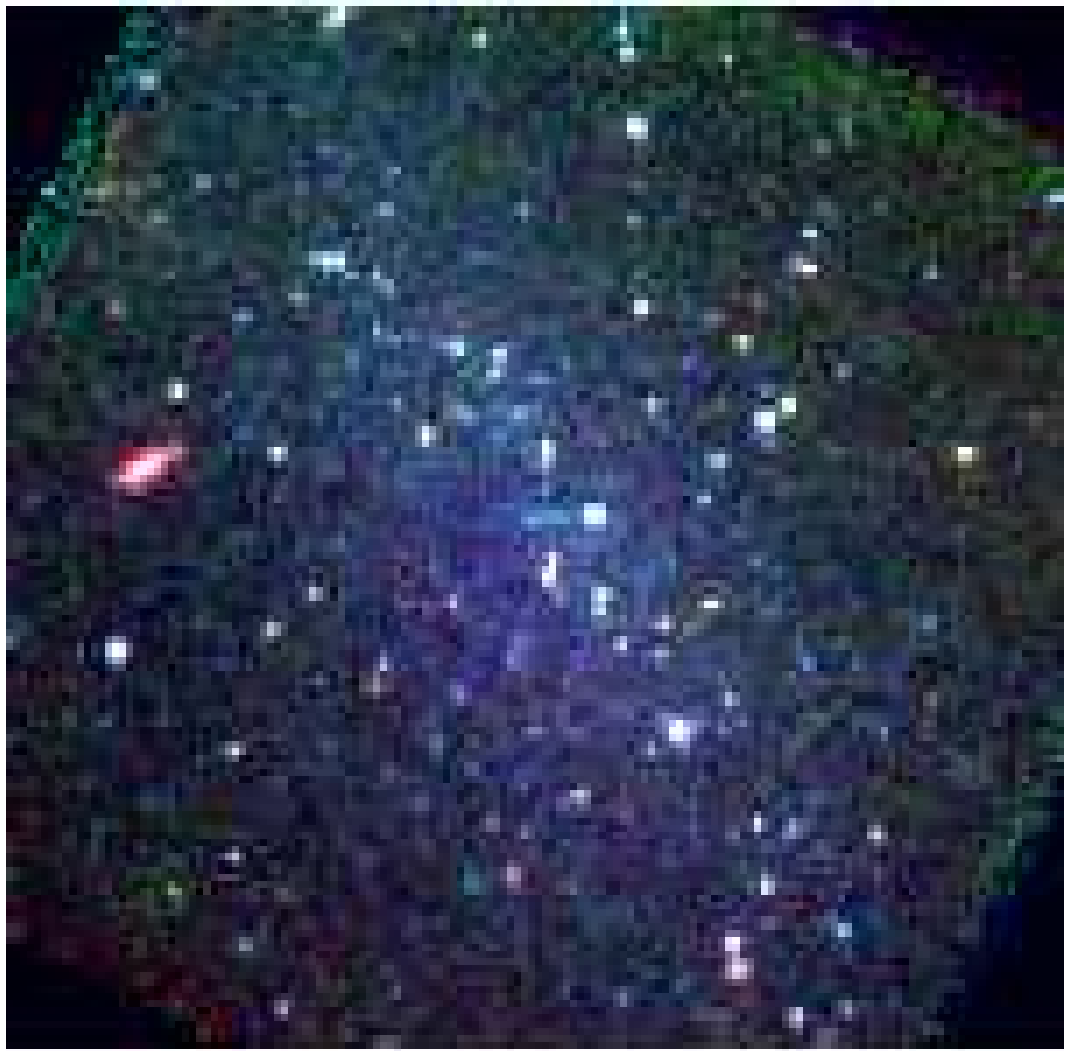}
}}
\caption{\label{fig:8} Three-color image of NGC~147 as seen by IRAC at
3.6 (blue), 5.8 (green) and 8\mum\ (red). The image shows that little
dust emission is present in this galaxy and that only a diffuse
stellar component is seen at shorter wavelength over a 5\arcmin\
square region, with N up and E to the left.}
\end{figure}

\newpage
\begin{figure}[h!]
\centerline{\hbox{
\includegraphics[width=400pt,height=400pt,angle=0]{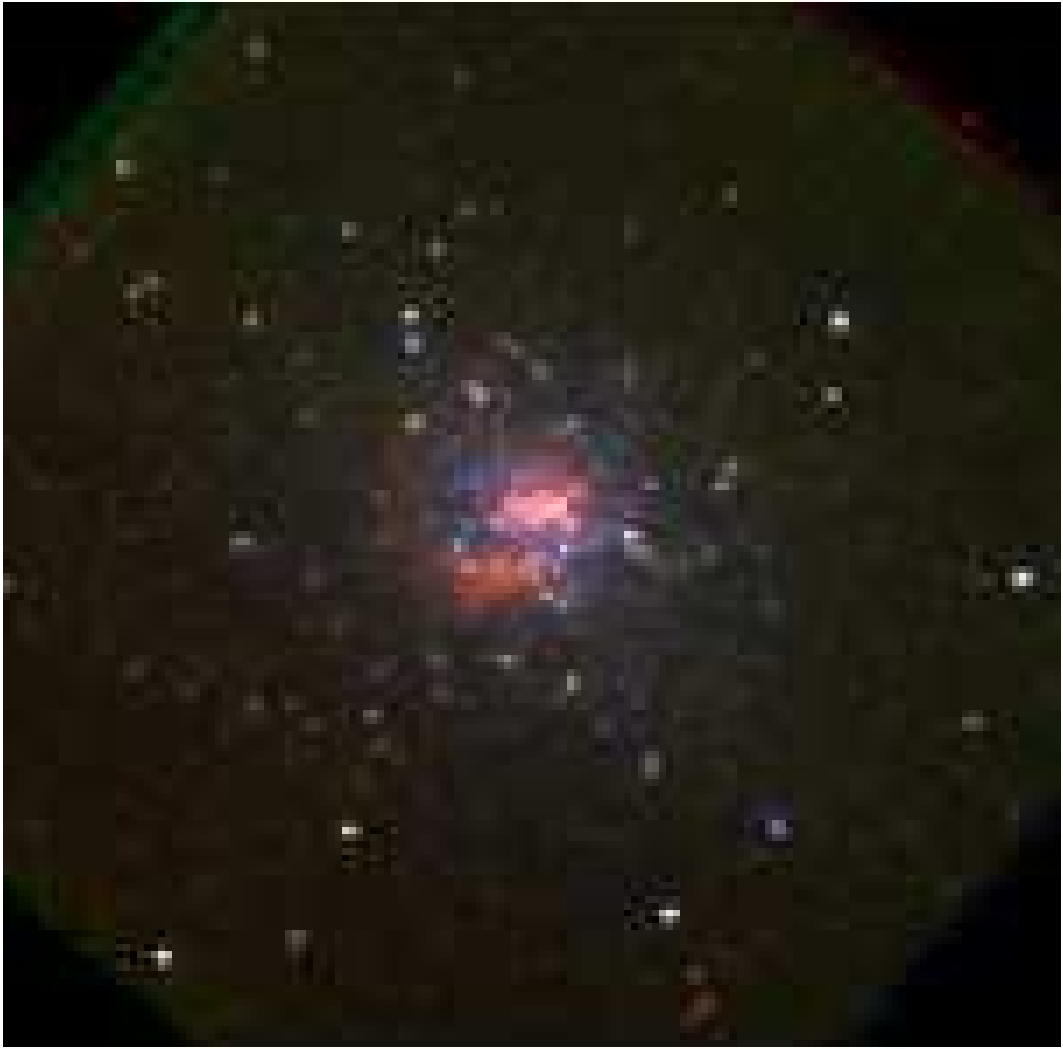}
}}
\caption{\label{fig:9} Three-color image of NGC~185 as seen by
IRAC at 3.6 (blue), 5.8 (green) and 8\mum\ (red). The image emphasizes
the dust clouds distribution seen at longer wavelengths over a 5\arcmin\
square region, with N up and E to the left.} 
\end{figure}

\newpage
\begin{figure}[h!]
\centerline{\hbox{
\includegraphics[width=400pt,height=400pt,angle=0]{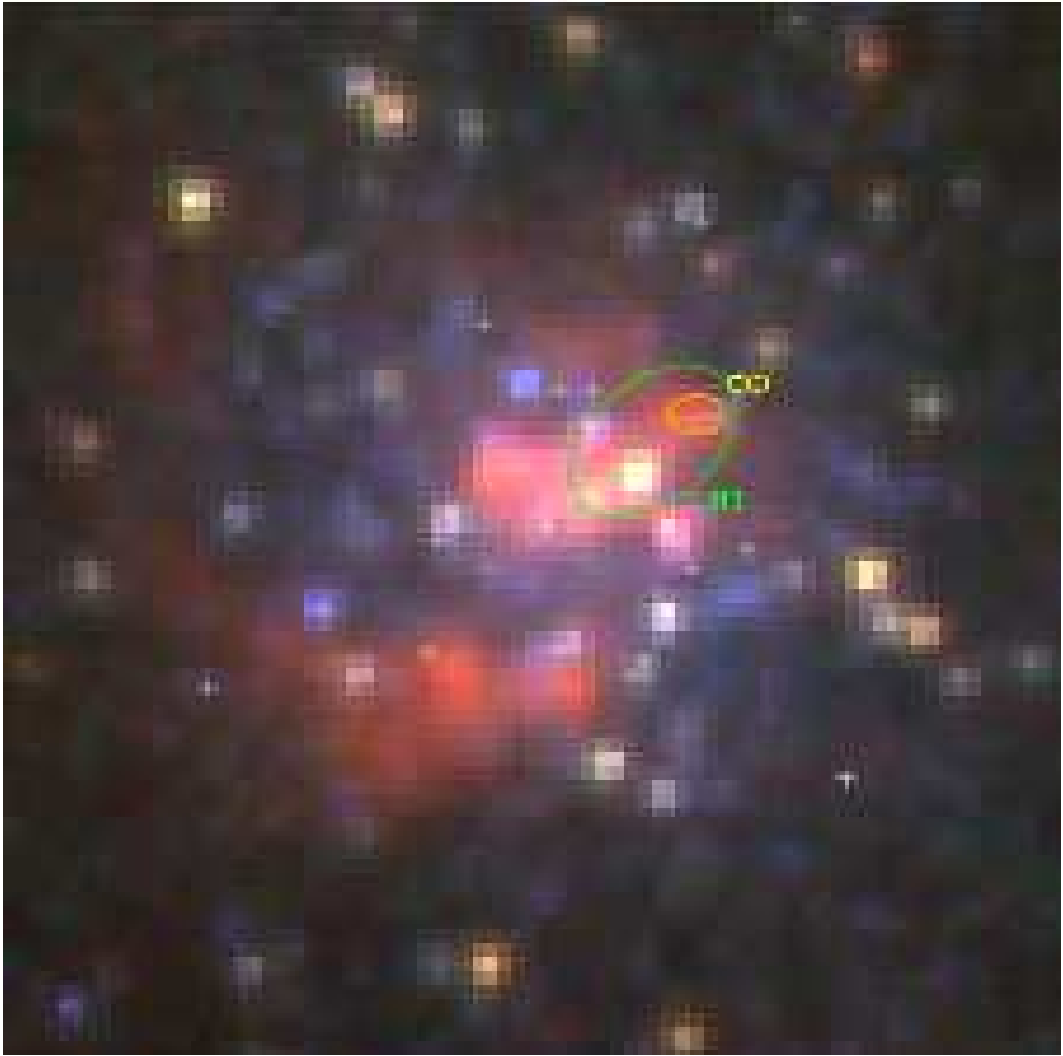}
}}
\caption{\label{fig:10} First main peak of each of the CO ({\it yellow
contour}) and HI emission ({\it green contour}) (Young 2001) overlayed
on the IRAC three-color image of NGC~185. The white plus signs
indicate the locations of ``Baade's blue stars,'' fifteen bright blue
objects with a minimum age of 100~Myr originally reported by Baade
(1951) and studied in more detail by Lee, Freedman, \& Madore (1993)
and Martínez-Delgado, Aparicio \& Gallart (1999). The image covers a
1\arcmin\ 30\arcsec\ square region centered on NGC~185, with N up and
E to the left.}
\end{figure}

\newpage
\begin{figure}[h!]
\centerline{\hbox{
\includegraphics[width=400pt,height=400pt,angle=0]{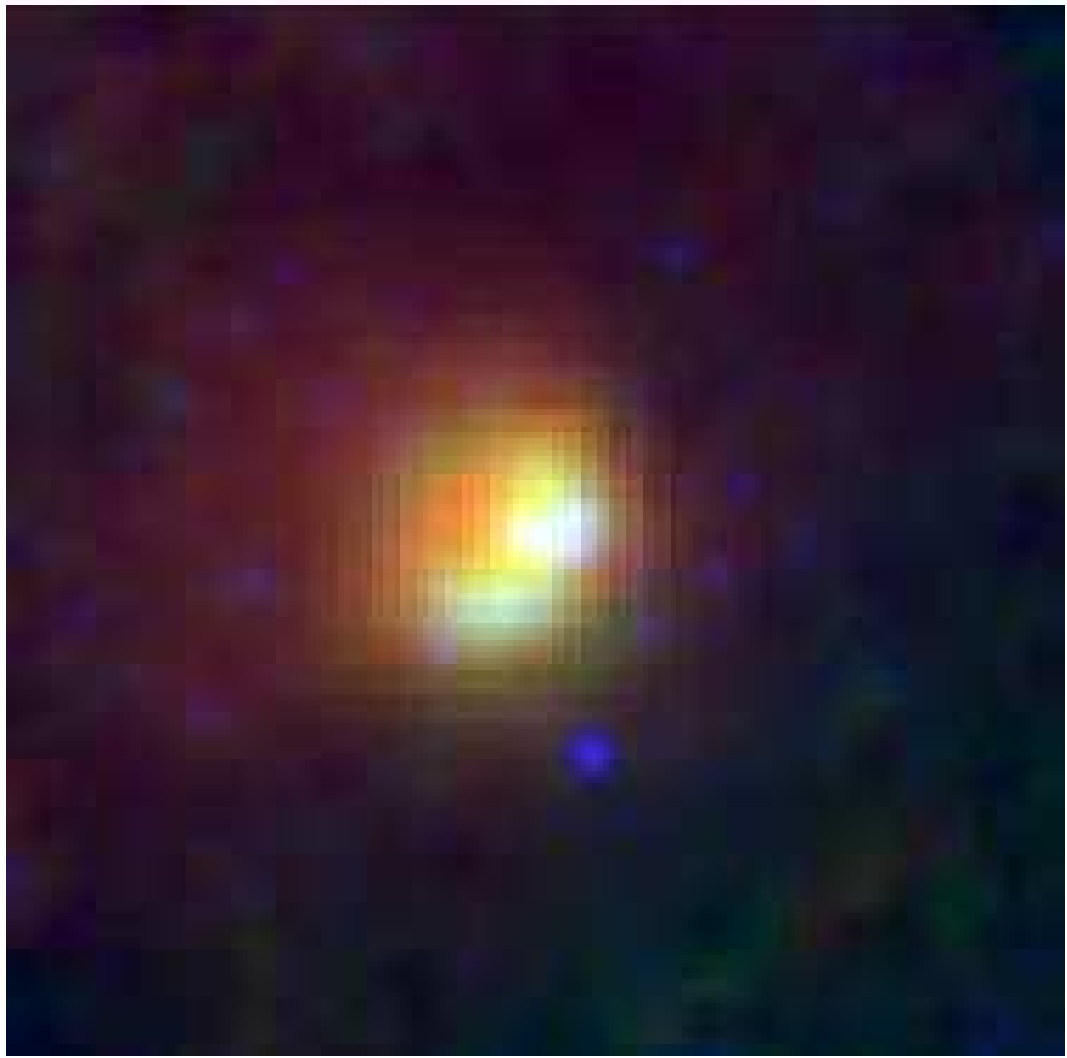}
}}
\caption{\label{fig:11} Three-color image of NGC~185 as seen by MIPS
at 24 ({\it blue}), 70 ({\it green}) and 160\mum\ ({\it red}). The
image shows that emission coming from the cold dust ({\it red}) is
more extended that the one coming from other dust temperature and size
components. The FOV of this image is 5\arcmin\ $\times$ 5\arcmin\ with
N up and E to the left.}
\end{figure}

\newpage
\begin{figure}[h!]
\centerline{\hbox{
\includegraphics[width=400pt,height=534pt,angle=90]{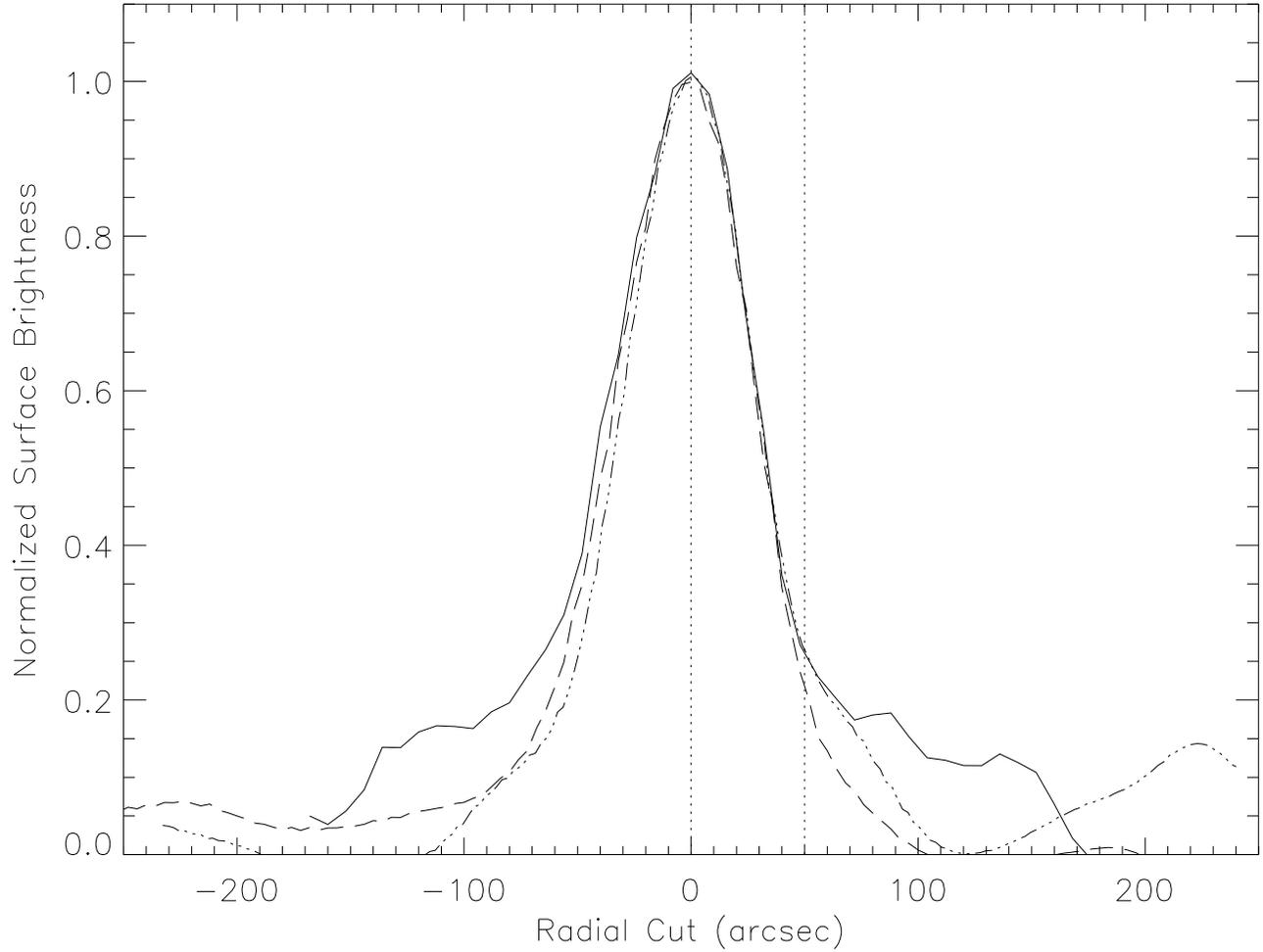}
}}
\caption{\label{fig:12} Comparison of the surface brightness radial
profiles of NGC~185 in the three MIPS bands: 160\mum\ ({\it solid
line}), 70\mum\ ({\it dashed line}) and 24\mum\ ({\it dash-dotted
line}). The three bands show similar profiles within the central
50\arcsec\ ({\it dotted line}) but beyond this radius, the 160\mum\ 
emission is more extended than that at 24 and 70\mum.}
\end{figure}

\newpage
\begin{figure}[h!]
\centerline{\includegraphics[width=240pt,height=185pt,angle=0]{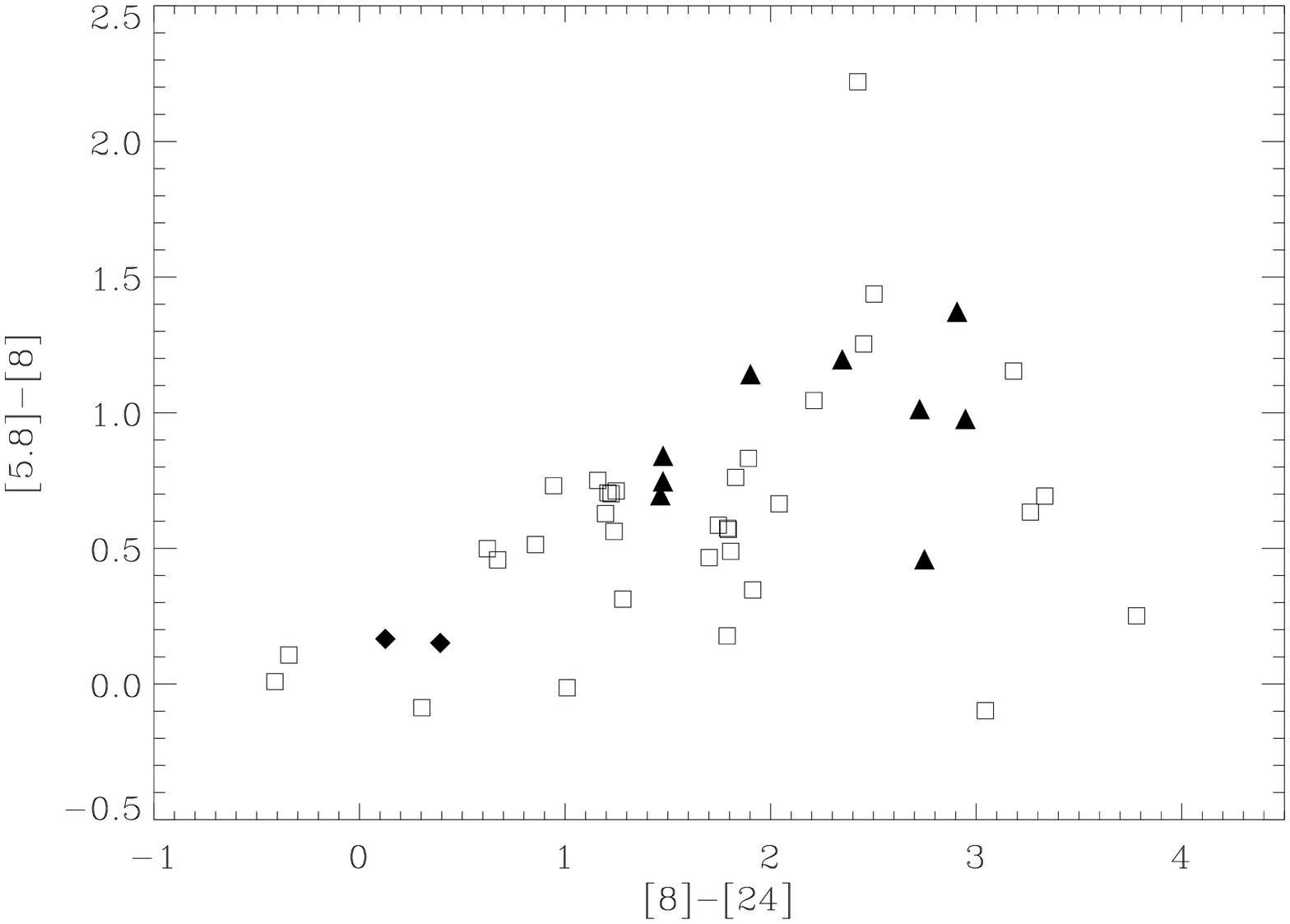}}
\centerline{\includegraphics[width=200pt,height=200pt,angle=0]{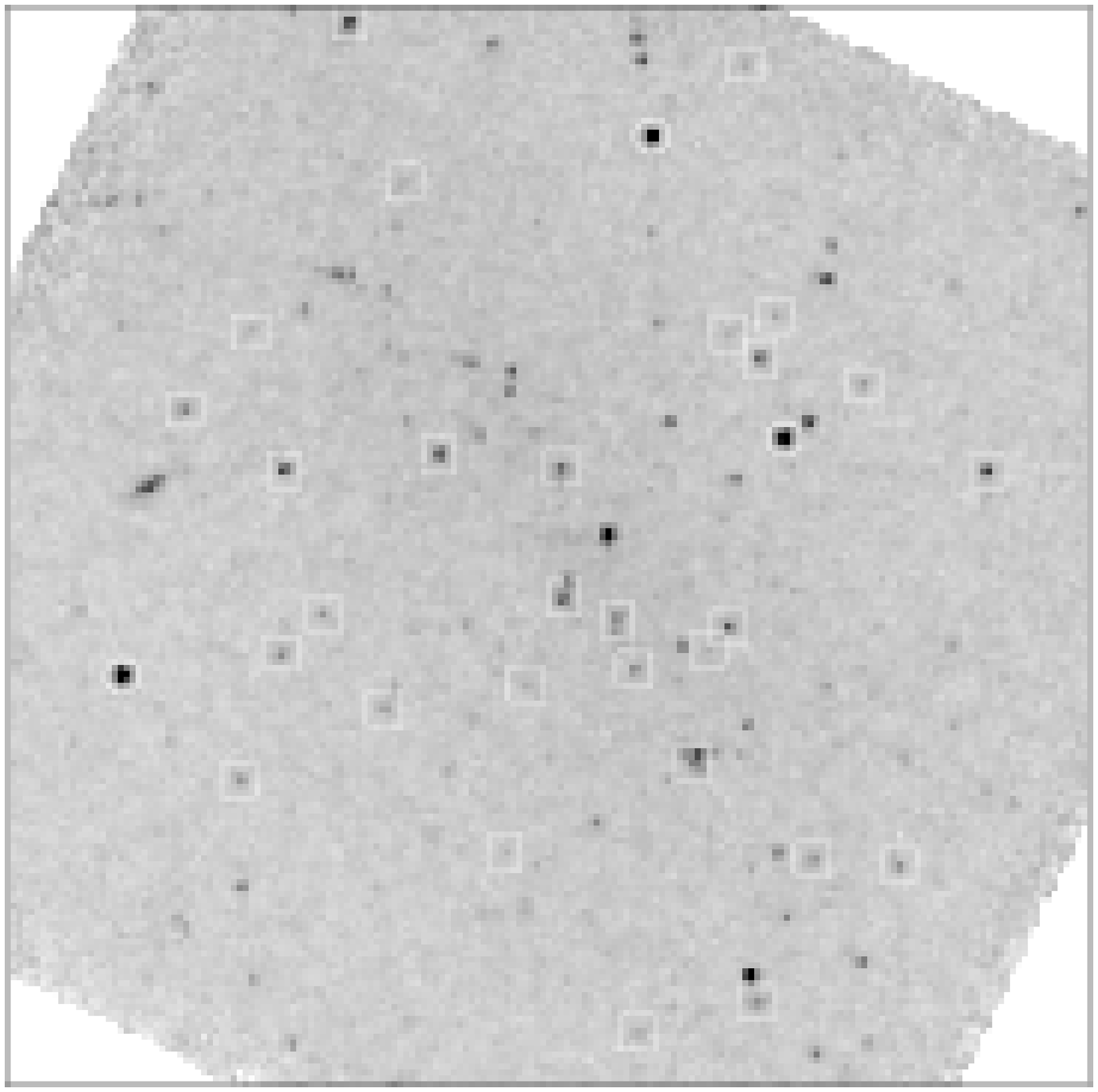}}
\caption{\label{fig:13} {\it Top:} Color-color diagram of a sample of
K Giant stars ({\it filled diamonds}), AGB stellar templates ({\it
filled triangles}), and the point sources extracted in the field of
NGC~147 observe with IRAC and MIPS ({\it open squares}). {\it Bottom:}
The point sources shown in the top figure, overlayed on the 5.8\mum\
IRAC image of size 5\arcmin\ $\times$ 5\arcmin\ with N up and E to the
left. The AGBs and K Giants, as classified in Table~2, are identified
with {\it squares} and {\it circles}, respectively.}
\end{figure}

\newpage
\begin{figure}[h!]
\centerline{\includegraphics[width=240pt,height=185pt,angle=0]{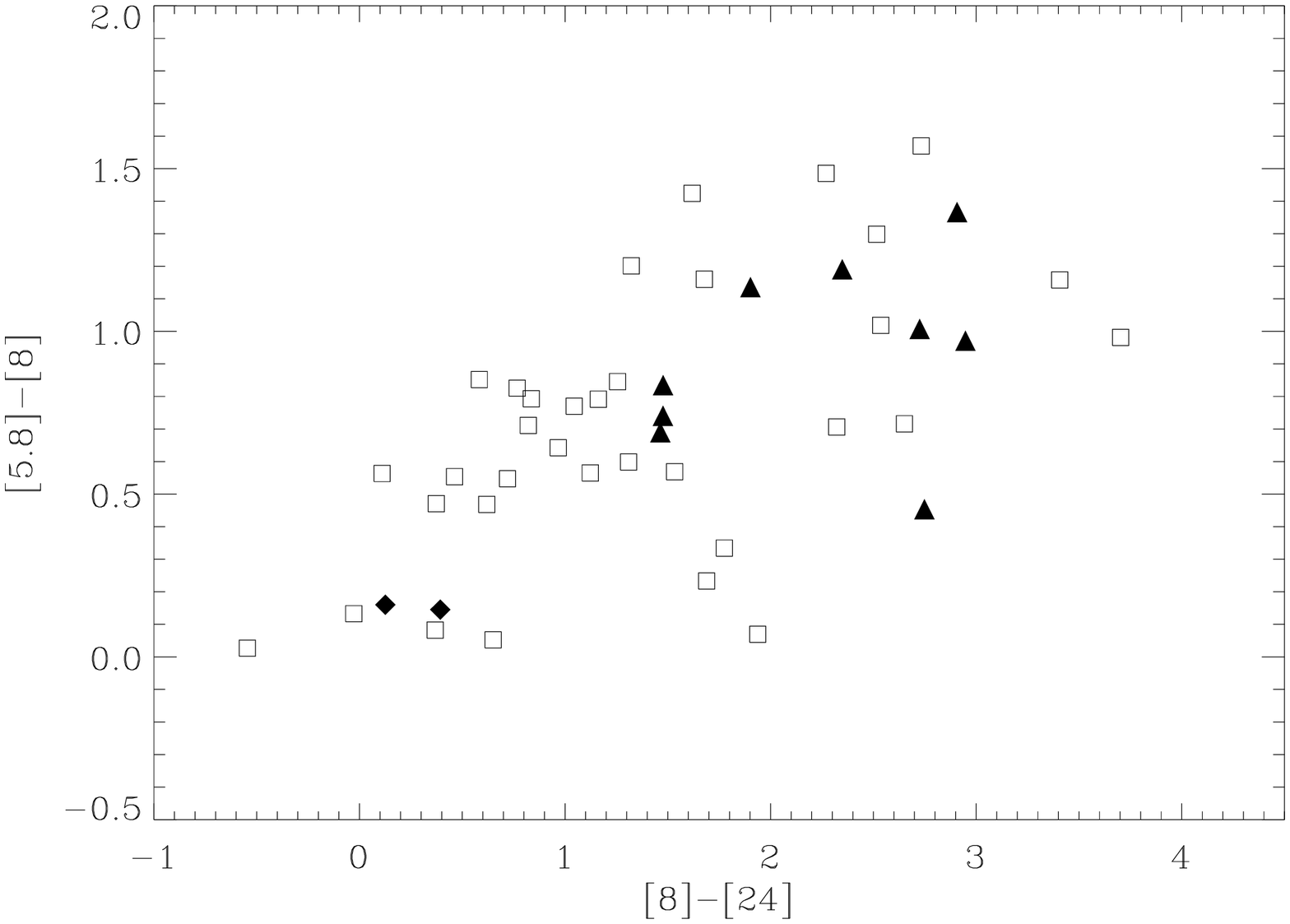}}
\centerline{\includegraphics[width=200pt,height=200pt,angle=0]{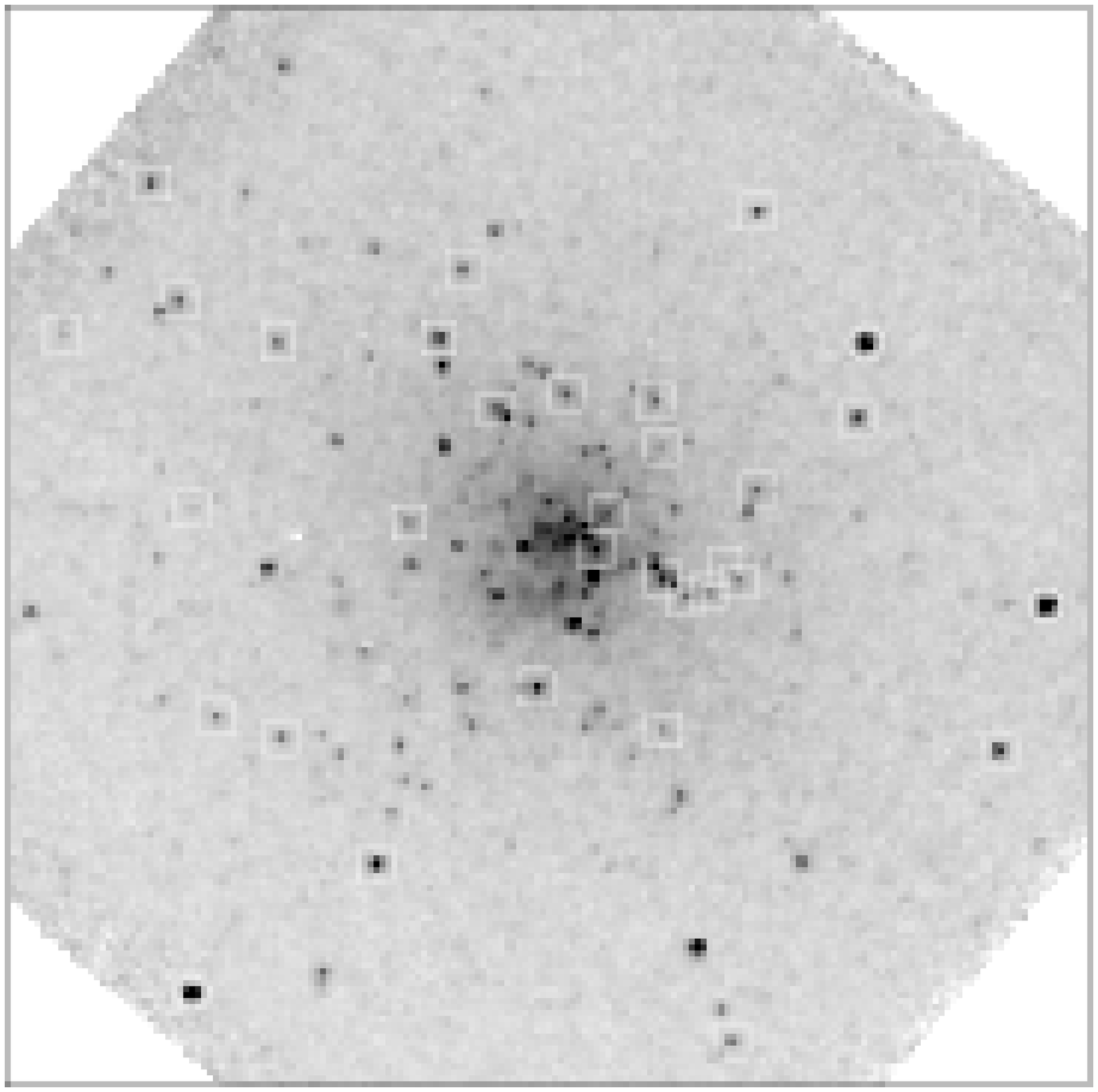}}
\caption{\label{fig:14} Same as Figure~13 but for NGC~185.}
\end{figure}

\newpage
\begin{figure}[h!]
\centerline{\hbox{
\includegraphics[width=400pt,height=400pt,angle=0]{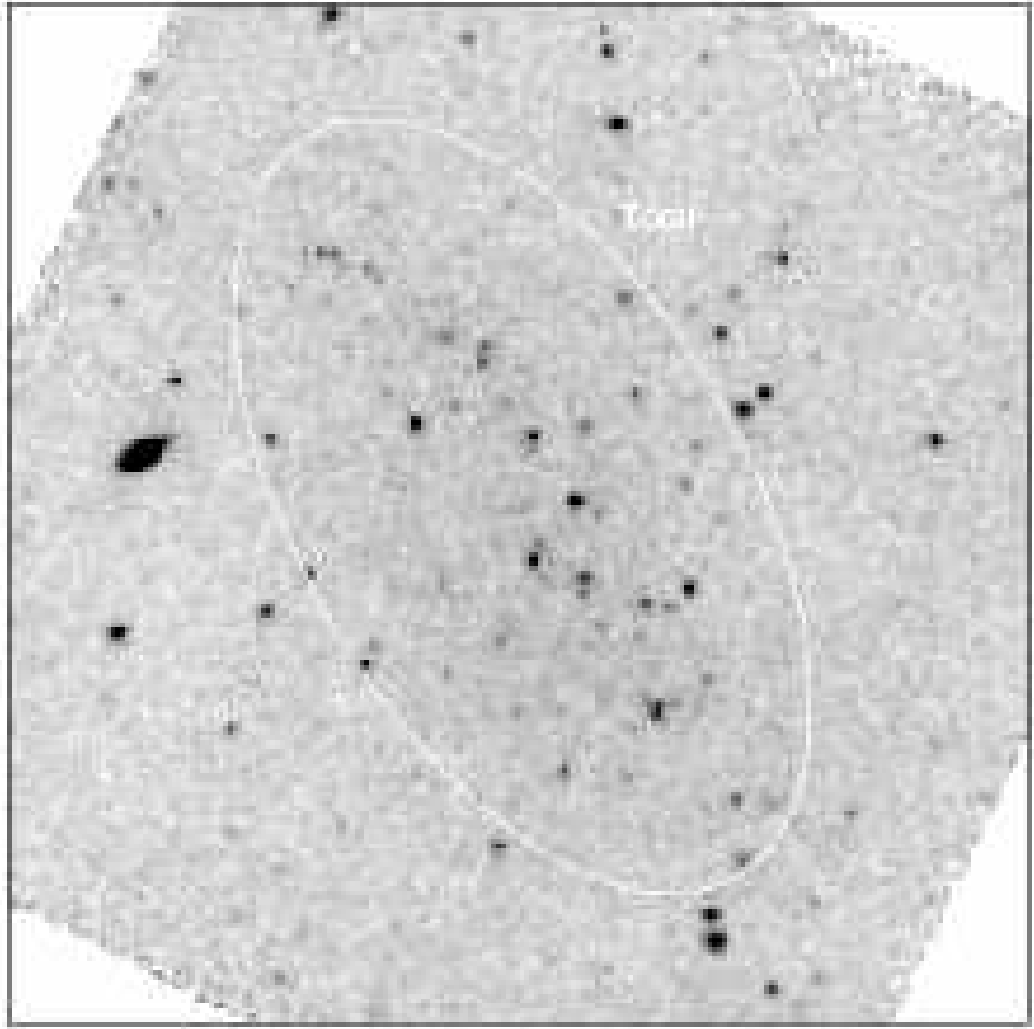}
}}
\caption{\label{fig:15} Emission region (as defined in Table~4), overlaid
on the IRAC 8\mum\ image of NGC~147. The FOV is 5\arcmin\ $\times$ 5\arcmin\ 
with N up and E to the left.}
\end{figure}

\newpage
\begin{figure}[h!]
\centerline{\hbox{
\includegraphics[width=400pt,height=400pt,angle=0]{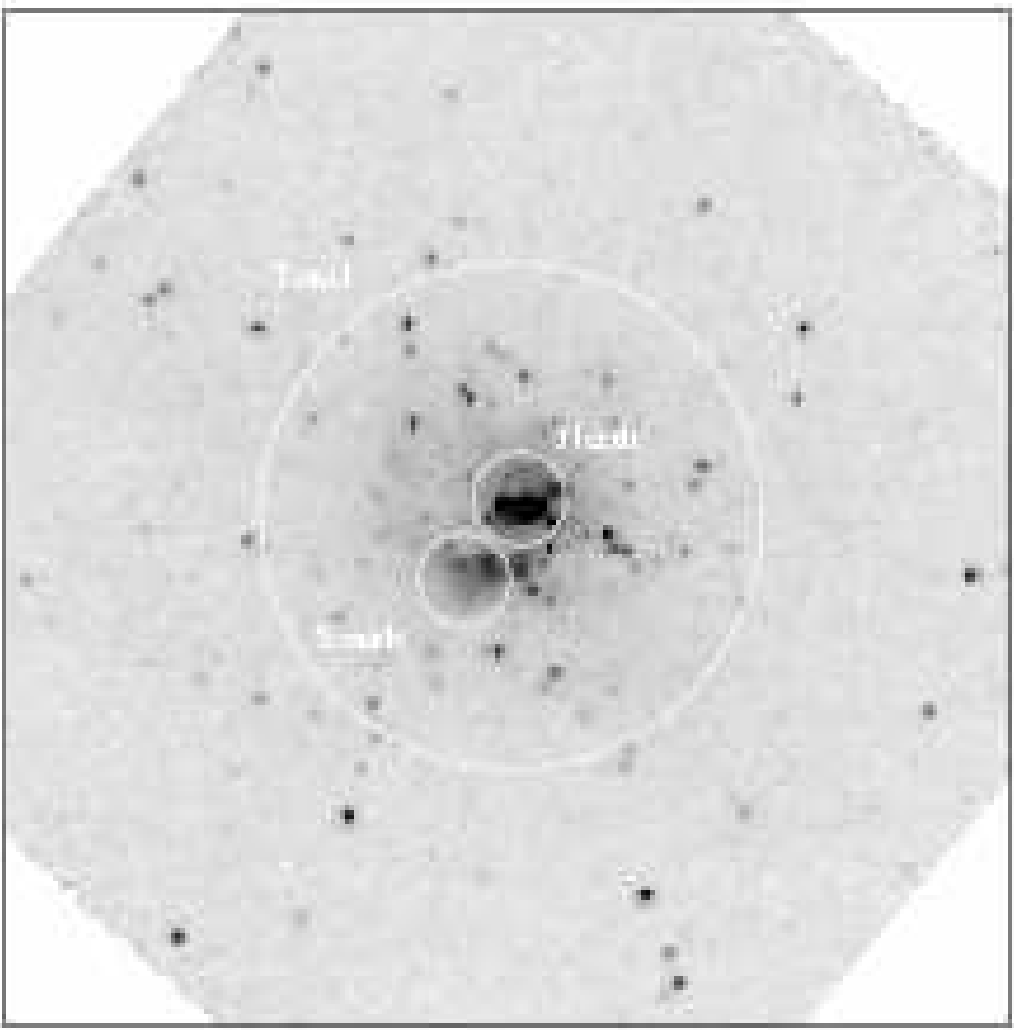}
}}
\caption{\label{fig:16} The two main emission regions, ``North'' and
``South'', as well as the ``Total'' region (as defined in Table~4),
overlaid on the IRAC 8\mum\ image of NGC~185. The FOV is 5\arcmin\
$\times$ 5\arcmin\ with N up and E to the left.}
\end{figure}

\newpage
\begin{figure}[h!]
\centerline{\hbox{
\includegraphics[width=300pt,height=400pt,angle=90]{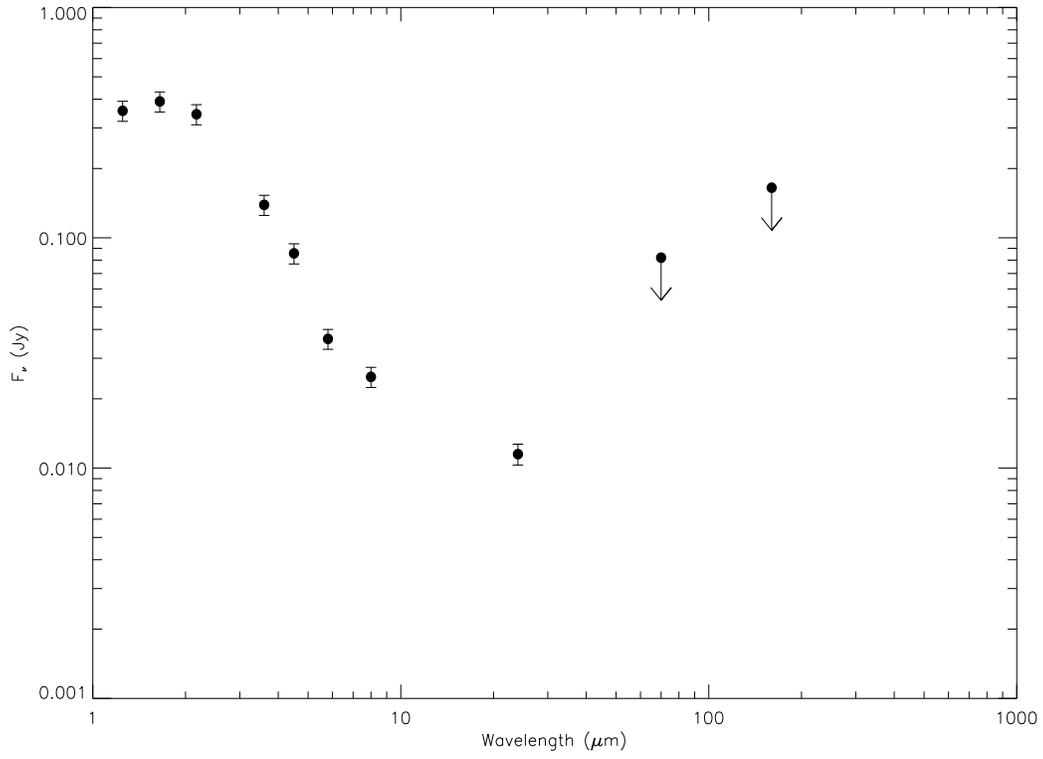}
}}
\caption{\label{fig:17} Infrared SED of ``NGC147Total'' region.
The {\it solid} points are the data points as listed in Table~5.
The {\it downward arrows} represent upper limits at 70 and 160\mum.
Since only upper limits are available at 70 and 160\mum, we have
almost no constraints on the dust mass for NGC~147 from SED fitting
(therefore not shown here).}
\end{figure}

\newpage
\begin{figure}[h!]
\centerline{\hbox{
\includegraphics[width=150pt,height=200pt,angle=90]{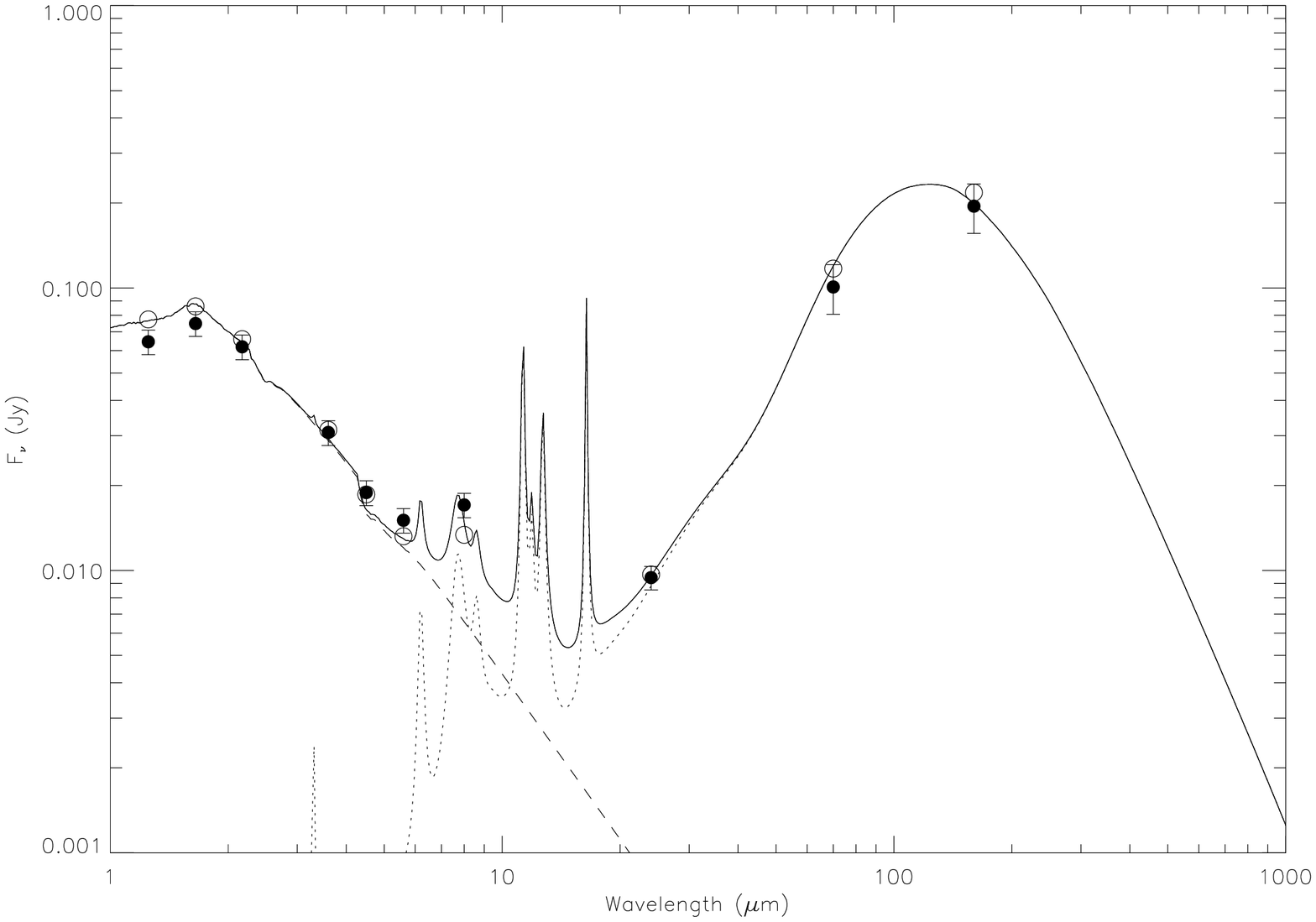}
}}
\centerline{\hbox{
\includegraphics[width=150pt,height=200pt,angle=90]{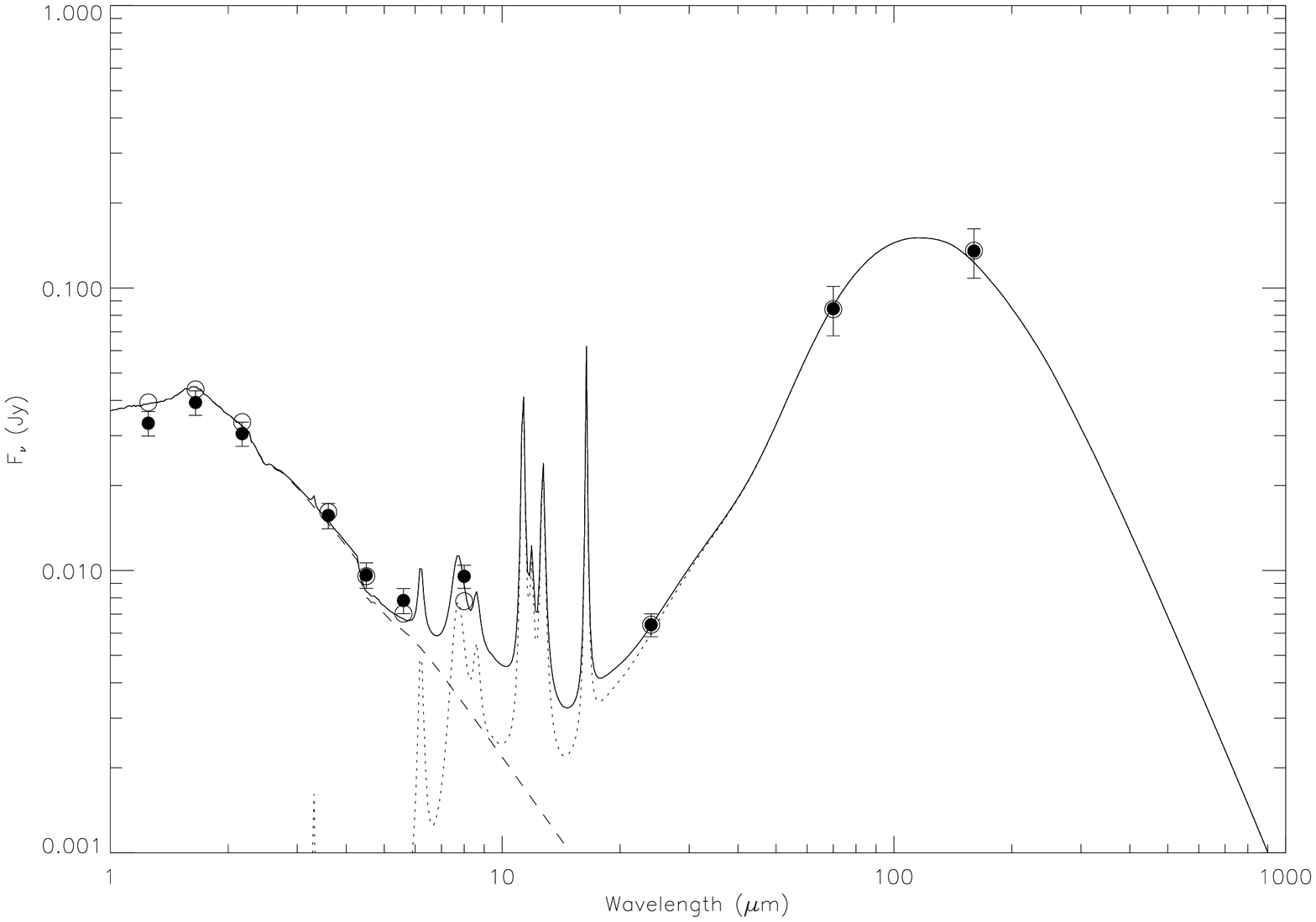}
}}
\centerline{\hbox{
\includegraphics[width=150pt,height=200pt,angle=90]{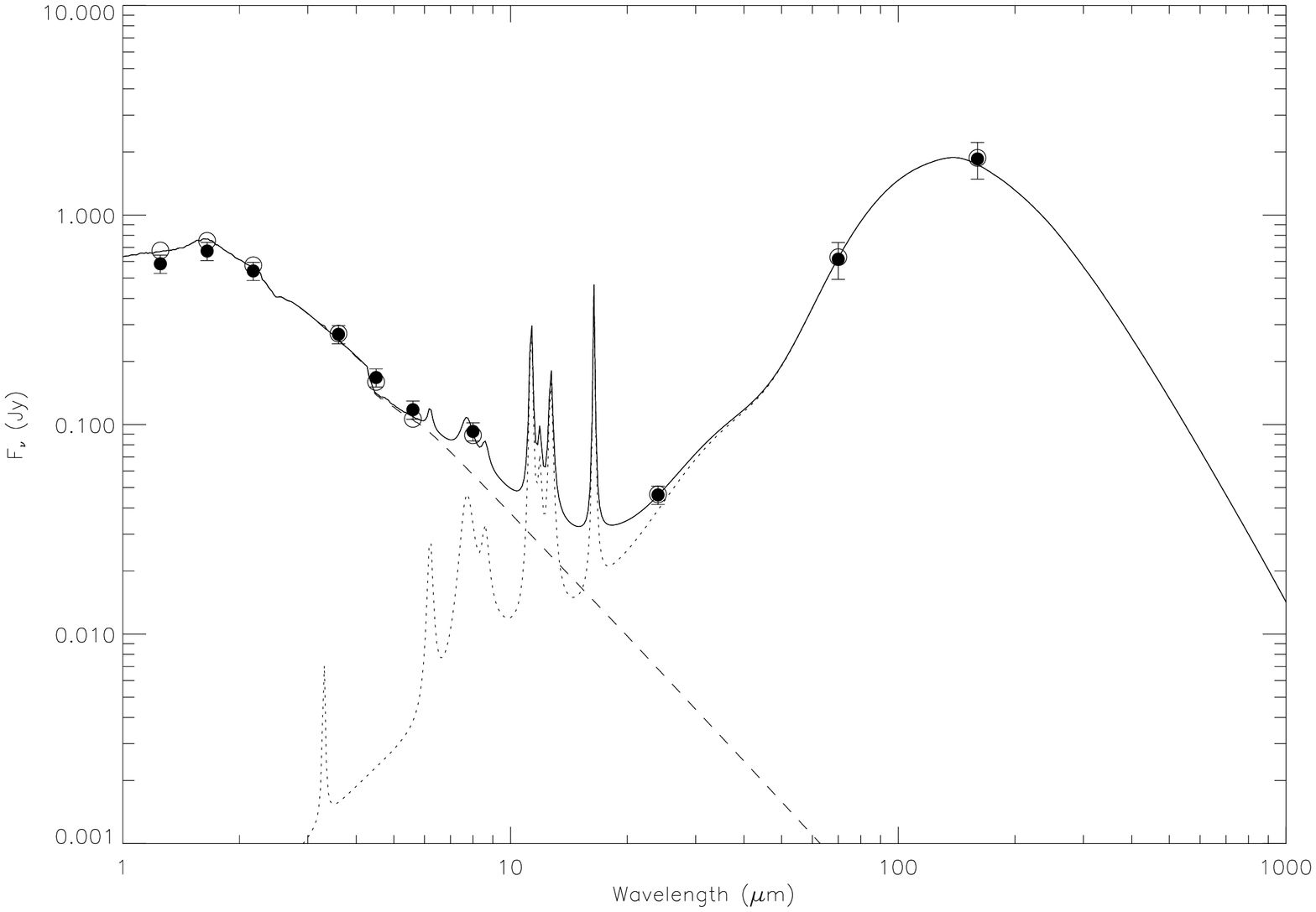}
}}
\caption{\label{fig:18} Infrared SED of the emission regions
``NGC185North'' ({\em top}), ``NGC185South'' ({\em center}) and
``NGC185Total'' ({\em bottom}). The {\it filled circles} are the data
and the {\it open circles} are the model multiplied with the passband
and integrated over wavelength. The {\it dashed line} is the stellar
component (PEGASE model), the {\it dotted line} is the dust component
(PAHs, graphites and silicates) and the {\it solid line} is the total
of all components. Note that in the case of the ``NGC185Total'',
the upper limit of the y-axis is larger by a factor of ten.}
\end{figure}

\newpage
\begin{figure}[h!]
\centerline{\hbox{
\includegraphics[width=400pt,height=534pt,angle=90]{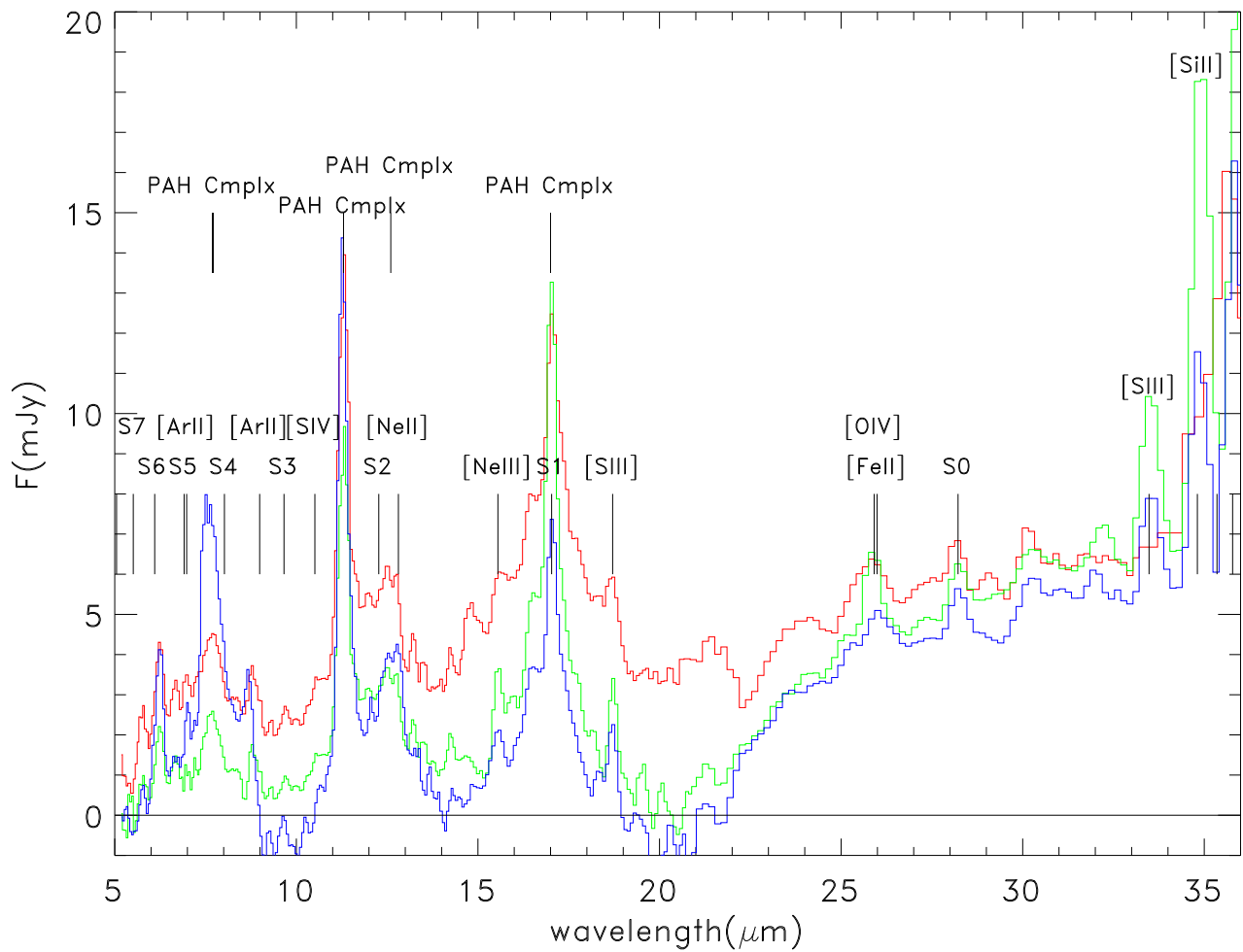}
}}
\caption{\label{fig:19} Extracted flux calibrated spectra for the
``NGC185IRSNorth'' ({\em red}), ``NGC185IRSCenter'' ({\em green}), and
``NGC185IRSSouth'' ({\em blue}) target positions. The main PAH
complexes at 6.2, 7.7, 11.3, 12.7 and 17\mum\ are indicated, as well
as most of the measured emission lines, including those from molecular
hydrogen. Deep silicate absorption features at $\sim$~9.7 and 18\mum\
are clearly seen at the ``NGC185IRSCenter'' and ``NGC185IRSSouth''
positions. The spectral properties resemble overall those of a
photodissociation region (except for the possible presence of [O~IV]
at 25.9\mum) and are similar to those displayed by starburst galaxies
or PAH-dominated Seyfert galaxies.}
\end{figure}

\newpage
\begin{figure}[h!]
\centerline{\hbox{
\includegraphics[width=150pt,height=400pt,angle=-90]{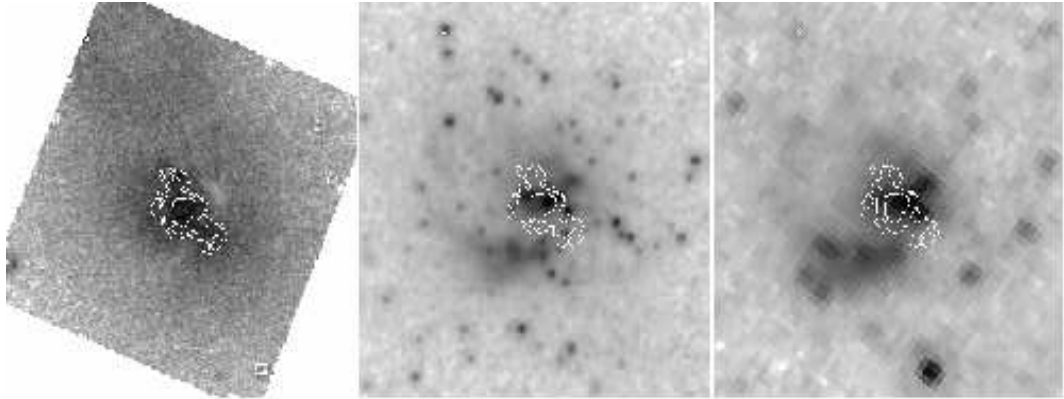}
}}
\caption{\label{fig:20} A comparison of the XMM-OM UVW1 observation
({\it left}) with those of {\it Spitzer} at 8\mum\ ({\it center}) and
24\mum\ ({\it right}) covering the central two arcminutes of NGC~185
with N up and E to the left. The grayscale is such that bright regions
are dark and extincted regions are white; the contours are those of
the UV diffuse emission.}
\end{figure}

\newpage
\begin{figure}[h!]
\centerline{\hbox{
\includegraphics[width=400pt,height=534pt,angle=90]{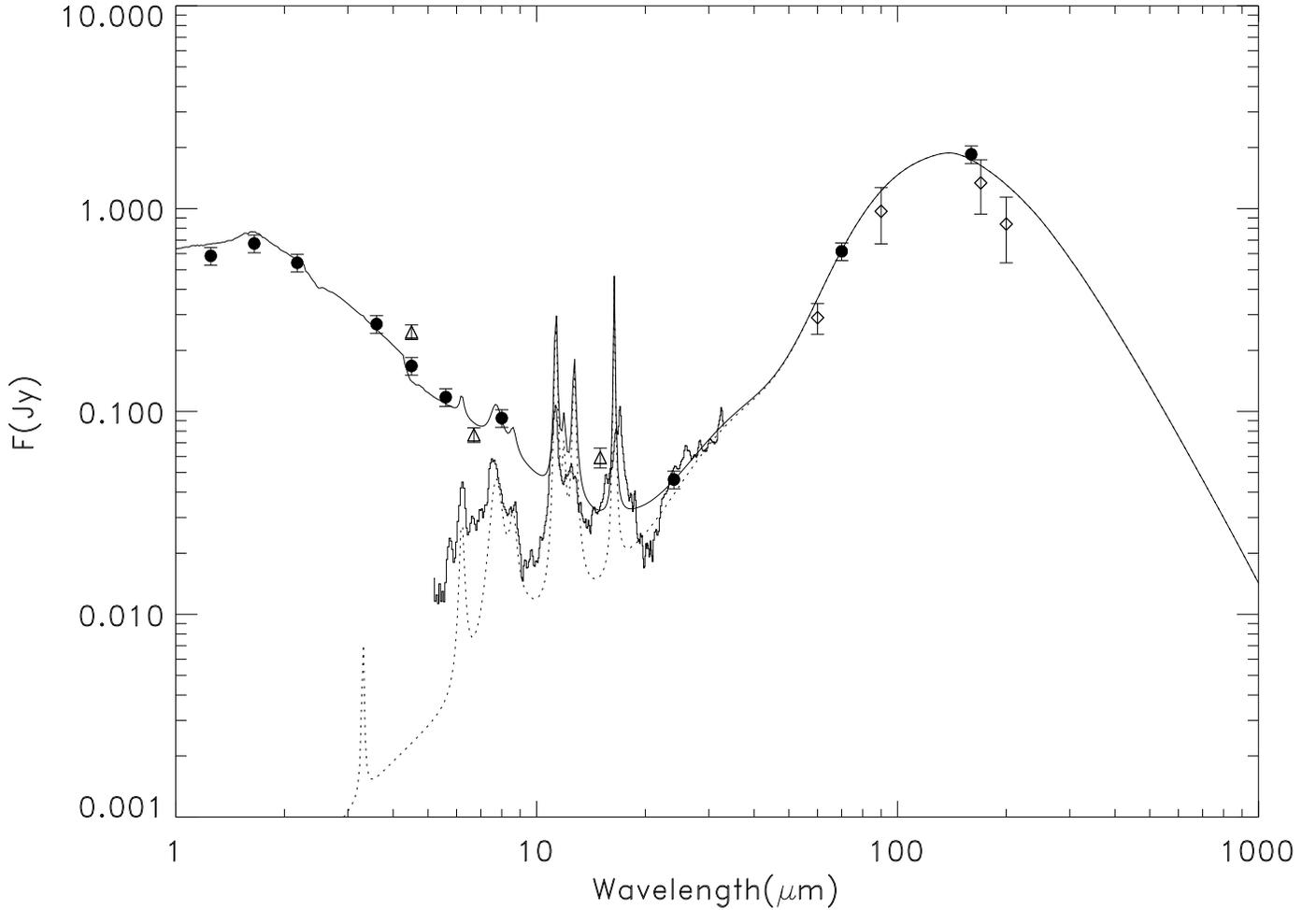}
}}
\caption{\label{fig:21} The average of the three IRS spectra overlayed
on the model SED for ``NGC185Total'' (see Figure~18). The average
spectrum was normalized to match the SED model in the $20-30$\mum\
wavelength region, where the stellar photometric contribution is the
lowest. The {\it dotted line} is the PAHs component of the SED
fit. The measurements shown on the plot are our photometry from IRAC
and MIPS ( {\it filled circles}) as well as the ISOCAM ({\it open
triangles}) and ISOPHOT ({\it open diamonds}) data points.}
\end{figure}

\newpage

\begin{table}[h!]
\centerline{TABLE 1}
\centerline{NGC~185 Spectral Extraction Positions \label{tbl-1}}
\vspace{0.2cm}
\begin{center}
\begin{tabular}{ccc}
\tableline\tableline
Position\tablenotemark{a}        &RA (J2000)   &Dec (J2000) \\
            & $^h$~~$^m$~~$^s$ & \arcdeg~~~ \arcmin~~~ \arcsec\ \\
\tableline
NGC185IRSNorth                    &0 38 56.57   &48 20 22.7 \\ 
NGC185IRSCenter                   &0 38 57.97   &48 20 14.6 \\ 
NGC185IRSSouth                    &0 38 57.63   &48 19 58.7 \\ 
NGC185IRSSky                      &0 38 29.93   &48 24 16.4 \\
\tableline
\end{tabular}
\tablenotetext{a}{Extraction position as shown in Figure~6 (except for
NGC185IRSSky which lies outside the field-of-view).}
\end{center}
\end{table}

\newpage
\begin{table}[h!]
\centerline{TABLE 2}
\centerline{Photometry of Infrared Point Sources in NGC~147 \label{tbl-2}}
\vspace{0.2cm}
\footnotesize
\begin{center}
\begin{tabular}{ccccccc}
\tableline\tableline
Source      &Type\tablenotemark{a} &RA (J2000)   &Dec (J2000)  &5.8\mum\  &8.0\mum\  &24.0\mum\ \\
            &     &$^h$~~$^m$~~$^s$ &\arcdeg~~~ \arcmin~~~ \arcsec\ & \multicolumn{3}{c}{Flux Density in mJy}\\
\tableline
    1 &AGB &0 33 12.83 &48 27 51.94   &0.672   &0.360   &0.098\\
    2 &AGB &0 33 11.78 &48 30 16.12   &0.662   &0.685   &0.227\\
    3 &AGB &0 33 15.25 &48 30 56.11   &0.617   &0.675   &0.391\\
    4 &AGB &0 33 17.76 &48 32 56.07   &0.581   &0.499   &0.095\\
    5 &AGB &0 32 59.94 &48 30 51.28   &0.556   &0.602   &0.188\\
    6 &AGB &0 33 19.56 &48 30 51.65   &0.496   &0.432   &0.102\\
    7 &AGB &0 33 11.81 &48 30 52.36   &0.461   &0.538   &0.330\\
    8 &AGB &0 33 19.65 &48 30 00.81   &0.425   &0.387   &0.130\\
    9 &AGB &0 33 22.35 &48 31 08.59   &0.395   &0.410   &0.134\\
   10 &AGB &0 33 06.28 &48 31 22.84   &0.373   &0.347   &0.186\\
   11 &AGB &0 33 06.40 &48 28 24.10   &0.361   &1.513   &1.514\\
   12 &AGB &0 33 10.27 &48 30 10.56   &0.358   &0.296   &0.059\\
   13 &AGB &0 33 07.19 &48 30 07.87   &0.355   &0.611   &0.628\\
   14 &AGB &0 33 08.17 &48 29 30.23   &0.344   &0.249   &0.087\\
   15 &AGB &0 33 02.43 &48 29 02.06   &0.315   &0.290   &0.162\\
   16 &AGB &0 33 20.80 &48 29 25.98   &0.307   &0.297   &0.096\\
   17 &AGB &0 33 18.45 &48 30 11.55   &0.268   &0.285   &0.073\\
   18 &AGB &0 33 09.80 &48 29 56.23   &0.223   &0.233   &0.079\\
   19 &AGB &0 33 04.83 &48 29 03.85   &0.222   &0.110   &0.195\\
   20 &AGB &0 33 20.53 &48 31 29.89   &0.191   &0.122   &0.068\\
   21 &AGB &0 33 16.77 &48 29 45.21   &0.179   &0.365   &0.393\\
   22 &AGB &0 33 03.43 &48 31 15.05   &0.168   &0.143   &0.081\\
   23 &AGB &0 33 06.70 &48 32 44.11   &0.168   &0.140   &0.072\\
   24 &AGB &0 33 09.69 &48 28 15.59   &0.155   &0.106   &0.370\\
   25 &AGB &0 33 16.24 &48 32 11.51   &0.150   &0.112   &0.070\\
   26 &AGB &0 33 05.87 &48 31 34.59   &0.149   &0.153   &0.354\\
   27 &AGB &0 33 07.14 &48 31 29.69   &0.144   &0.132   &0.074\\
   28 &AGB &0 33 07.75 &48 30 01.84   &0.119   &0.169   &0.139\\
   29 &AGB &0 33 13.43 &48 29 05.85   &0.118   &0.118   &0.083\\
   30 &AGB &0 33 12.82 &48 29 52.15   &0.109   &0.106   &0.230\\
   31 &AGB &0 33 15.24 &48 33 17.58   &0.107   &0.168   &0.338\\
   32 &KGiant &0 33 24.09 &48 29 54.60   &1.666   &0.997   &0.078\\
   33 &KGiant &0 33 09.32 &48 32 24.31   &1.646   &0.824   &0.117\\
   34 &KGiant &0 33 05.60 &48 31 00.34   &1.567   &0.857   &0.063\\
\tableline
\end{tabular}
\tablenotetext{a}{Classification based on their infrared colors.}
\end{center}
\end{table}

\newpage
\begin{table}[h!]
\centerline{TABLE 3}
\centerline{Photometry of Infrared Point Sources in NGC~185 \label{tbl-3}}
\vspace{0.2cm}
\footnotesize
\begin{center}
\begin{tabular}{ccccccc}
\tableline\tableline
Source      &Type\tablenotemark{a} &RA (J2000)   &Dec (J2000)  &5.8\mum\  &8.0\mum\  &24.0\mum\ \\
            &     &$^h$~~$^m$~~$^s$ &\arcdeg~~~ \arcmin~~~ \arcsec\ & \multicolumn{3}{c}{Flux Density in mJy}\\
\tableline
    1 &AGB &0 39 02.86 &48 18 45.59   &1.003   &1.198   &0.220\\
    2 &AGB &0 39 01.10 &48 21 11.52   &0.838   &0.879   &0.201\\
    3 &AGB &0 38 58.39 &48 19 34.44   &0.694   &0.809   &0.176\\
    4 &AGB &0 38 45.53 &48 19 16.92   &0.616   &0.564   &0.067\\
    5 &AGB &0 38 56.75 &48 20 12.55   &0.540   &1.156   &1.004\\
    6 &AGB &0 39 08.34 &48 21 21.79   &0.534   &0.449   &0.068\\
    7 &AGB &0 39 09.11 &48 21 54.37   &0.532   &0.480   &0.100\\
    8 &AGB &0 38 49.45 &48 20 49.24   &0.528   &0.520   &0.136\\
    9 &AGB &0 38 52.24 &48 21 46.49   &0.521   &0.437   &0.083\\
   10 &AGB &0 38 54.91 &48 20 04.72   &0.513   &0.472   &0.208\\
   11 &AGB &0 39 05.58 &48 21 10.26   &0.480   &0.791   &0.287\\
   12 &AGB &0 39 00.42 &48 21 30.66   &0.476   &0.432   &0.071\\
   13 &AGB &0 38 59.50 &48 20 51.81   &0.463   &0.438   &0.157\\
   14 &AGB &0 38 57.58 &48 20 56.21   &0.459   &0.545   &0.186\\
   15 &AGB &0 38 52.78 &48 20 03.94   &0.437   &0.484   &0.136\\
   16 &AGB &0 38 52.27 &48 20 29.22   &0.389   &0.787   &0.375\\
   17 &AGB &0 39 05.50 &48 19 20.62   &0.389   &0.440   &0.102\\
   18 &AGB &0 38 52.98 &48 17 56.38   &0.354   &0.819   &1.089\\
   19 &AGB &0 38 53.63 &48 20 01.15   &0.294   &0.218   &0.120\\
   20 &AGB &0 38 54.26 &48 19 59.38   &0.289   &0.265   &0.080\\
   21 &AGB &0 38 55.09 &48 20 53.72   &0.278   &0.441   &0.222\\
   22 &AGB &0 39 01.93 &48 20 20.29   &0.277   &0.313   &0.098\\
   23 &AGB &0 39 07.27 &48 19 26.24   &0.262   &0.177   &0.090\\
   24 &AGB &0 38 54.86 &48 19 22.76   &0.253   &0.147   &0.094\\
   25 &AGB &0 38 56.44 &48 20 22.33   &0.240   &0.380   &0.940\\
   26 &AGB &0 38 54.95 &48 20 41.44   &0.225   &0.235   &0.214\\
   27 &AGB &0 39 16.96 &48 19 36.79   &0.192   &0.346   &0.377\\
   28 &AGB &0 39 11.55 &48 21 12.41   &0.130   &0.137   &0.169\\
   29 &AGB &0 38 53.18 &48 20 07.76   &0.107   &0.144   &0.468\\
   30 &AGB &0 39 07.94 &48 20 23.90   &0.079   &0.110   &0.122\\
   31 &KGiant &0 38 49.23 &48 21 10.08   &1.901   &1.062   &0.069\\
   32 &KGiant &0 38 44.23 &48 19 57.22   &1.754   &1.080   &0.113\\
   33 &KGiant &0 39 07.97 &48 18 10.01   &1.580   &0.929   &0.140\\
   34 &KGiant &0 38 41.00 &48 21 05.91   &1.280   &0.732   &0.143\\
\tableline
\end{tabular}
\tablenotetext{a}{Classification based on their infrared colors.}
\end{center}
\end{table}

\newpage
\begin{table}[h!]
\centerline{TABLE 4}
\centerline{NGC~147 and NGC~185 Emission Regions \label{tbl-4}}
\vspace{0.2cm}
\begin{center}
\begin{tabular}{cccc}
\tableline\tableline
Region\tablenotemark{a}        &RA (J2000)   &Dec (J2000)  &Area \\
            &$^h$~~$^m$~~$^s$ &\arcdeg~~~ \arcmin~~~ \arcsec\ &\arcmin$^2$ \\
\tableline
NGC185North                    &0 38 57.73  &48 20 19.7   &0.17 \\
NGC185South                    &0 38 59.39  &48 19 55.7   &0.17 \\
NGC185Total                    &0 38 57.97  &48 20 14.6   &4.91 \\
NGC147Total                    &0 33 12.12  &48 30 31.5   &7.26 \\
\tableline
\end{tabular}
\tablenotetext{a}{Circular/elliptical aperture as shown in Figure~15 and 16.}
\end{center}
\end{table}

\newpage
\begin{table}[h!]
\centerline{TABLE 5}
\centerline{Photometry of Emission Regions in NGC~147 \label{tbl-5}}
\vspace{0.2cm}
\begin{center}
\begin{tabular}{cc}
\tableline\tableline
Wavelength & Total \\
\mum     & \multicolumn{1}{c}{Flux Density in mJy}\\
\tableline
1.22 & 356 $\pm$ 36\\
1.65 & 391 $\pm$ 39\\
2.16 & 344 $\pm$ 34\\
3.6  & 139 $\pm$ 14\\
4.5  &  86 $\pm$ 9\\
5.8  &  51 $\pm$ 5\\
8    &  34 $\pm$ 3\\
24   &  21 $\pm$ 2\\
70   &  82\tablenotemark{a} $\pm$ 17\\
160  & 165\tablenotemark{a} $\pm$ 32\\
\tableline
\end{tabular}
\tablenotetext{a}{1$\sigma$ upper limit.}
\end{center}
\end{table}

\newpage
\begin{table}[h!]
\centerline{TABLE 6}
\centerline{Photometry of Emission Regions in NGC~185 \label{tbl-6}}
\vspace{0.2cm}
\begin{center}
\begin{tabular}{cccc}
\tableline\tableline
Wavelength & North & South & Total \\
\mum     & \multicolumn{3}{c}{Flux Density in mJy}\\
\tableline
1.22 &  64 $\pm$ 6   &  33 $\pm$ 3   & 583 $\pm$ 58\\
1.65 &  75 $\pm$ 8   &  39 $\pm$ 4   & 671 $\pm$ 67\\
2.16 &  62 $\pm$ 6   &  30 $\pm$ 3   & 539 $\pm$ 54\\
3.6  &  31 $\pm$ 3   &  16 $\pm$ 2   & 269 $\pm$ 27\\
4.5  &  19 $\pm$ 2   &  10 $\pm$ 1   & 167 $\pm$ 17\\
5.8  &  15 $\pm$ 2   &   8 $\pm$ 1   & 117 $\pm$ 12\\
8    &  17 $\pm$ 2   &  10 $\pm$ 1   &  92 $\pm$ 9\\
24   &   9 $\pm$ 1   &   6 $\pm$ 1   &  46 $\pm$ 5\\
70   & 101 $\pm$ 20  &  84 $\pm$ 17  & 614 $\pm$ 123\\
160  & 194 $\pm$ 39  & 134 $\pm$ 27  &1846 $\pm$ 369\\
\tableline
\end{tabular}
\end{center}
\end{table}

\newpage
\begin{table}[h!]
\centerline{TABLE 7}
\centerline{Stellar Photospheric and Dust Flux Density Contributions for NGC~185 \label{tbl-7}}
\vspace{0.2cm}
\footnotesize
\begin{center}
\begin{tabular}{llcccc}
\tableline\tableline
Region  &Wavelength  &Total\tablenotemark{a}  &Dust &Stellar\tablenotemark{b} &Dust Fraction\\
             &\mum   &mJy     &mJy    &      & \\
\tableline
NGC185North  &3.6    &29.4    &0.4    &1.0   &0.01 \\
             &4.5    &16.2    &0.5    &0.54  &0.03 \\ 
             &5.8    &12.6    &1.1    &0.40  &0.09 \\
             &8.0    &13.4    &6.8    &0.22  &0.51 \\
             &24.0   &8.9     &8.1    &0.03  &0.91 \\
             &70.0   &106.8   &106.7  &0.00  &1.00 \\
             &160.0  &184.9   &184.9  &0.00  &1.00 \\
\tableline
NGC185South  &3.6    &15.0    &0.3    &1.0   &0.02 \\
             &4.5    &8.4     &0.4    &0.54  &0.05 \\
             &5.8    &6.7     &0.8    &0.40  &0.12 \\
             &8.0    &8.6     &5.2    &0.23  &0.61 \\
             &24.0   &6.4     &6.0    &0.03  &0.94 \\
             &70.0   &86.8    &86.7   &0.00  &1.00 \\
             &160.0  &121.9   &121.9  &0.00  &1.00 \\
\tableline
NGC185Total   &3.6    &255.2  &1.3    &1.0   &0.01 \\
              &4.5    &139.4  &2.0    &0.54  &0.01 \\
              &5.8    &104.8  &4.1    &0.40  &0.04 \\
              &8.0    &79.8   &22.4   &0.23  &0.28 \\
              &24.0   &45.1   &38.2   &0.03  &0.85 \\
              &70.0   &610.3  &609.4  &0.00  &1.00 \\
              &160.0  &1720.8 &1720.6 &0.00  &1.00 \\
\tableline
\end{tabular}
\tablenotetext{a}{The measured flux density (photospheric plus dust contributions).}
\tablenotetext{b}{Stellar flux density normalized with respect to that at 3.6\mum.}
\end{center}
\end{table}

\newpage
\begin{table}[h!]
\centerline{TABLE 8}
\centerline{Mass and Estimates from SED Fits for NGC~185 \label{tbl-8}}
\vspace{0.2cm}
\begin{center}
\begin{tabular}{lccc}
\tableline\tableline
Cloud   &Metallicity  &Age  &Dust Mass  \\
        &             &Myr  &M$_\odot$  \\
\tableline
\\
NGC185North   &-0.7  &$0^{+60}$  &$1.4_{-0.6}^{+1.2} \times 10^{2}$ \\
\\
\tableline
\\
NGC185South   &-0.7  &$4_{-4}^{+96}$  &$7.7_{-3.7}^{+5.9} \times 10^{1}$ \\
\\
\tableline
\\
NGC185Total   &-0.7  &$450_{-330}^{+550}$  &$1.9_{-0.9}^{+1.9} \times 10^{3}$ \\
\\
\tableline
\end{tabular}
\end{center}
\end{table}

\newpage
\begin{table}[h!]
\centerline{TABLE 9}
\centerline{Mass Estimates for Each Component for NGC~185 \label{tbl-9}}
\vspace{0.2cm}
\footnotesize
\begin{center}
\begin{tabular}{lccc}
\tableline\tableline
Region   &PAH   &Graphite   &Silicate \\
\tableline
NGC185North   &$7.8_{-3.9}^{+6.7} \rm{M}_\odot$   &$4.11_{-2.05}^{+3.52} \times 10^{1} \rm{M}_\odot$   &$9.91_{-4.56}^{7.81} \times 10^{1} \rm{M}_\odot$ \\
\tableline
NGC185South   &$4.3_{-2.1}^{+3.3} \rm{M}_\odot$   &$2.26_{-1.09}^{+1.73} \times 10^{1} \rm{M}_\odot$   &$5.01_{-2.41}^{+3.84} \times 10^{1} \rm{M}_\odot$ \\
\tableline
NGC185Total   &$1.05_{-0.49}^{+1.05} \times 10^{2} \rm{M}_\odot$   &$5.58_{-2.64}^{+5.58} \times 10^{2} \rm{M}_\odot$   &$1.236_{-0.585}^{+1.236} \times 10^{3} \rm{M}_\odot$ \\
\tableline
\end{tabular}
\end{center}
\end{table}

\newpage
\begin{table}[h!]
\centerline{TABLE 10}
\centerline{Emission Lines in NGC~185 (Corrected for Extinction) \label{tbl-10}}
\vspace{0.2cm}
\footnotesize
\begin{center}
\begin{tabular}{ccccc}
\tableline\tableline
Line &Wavelength &NGC185IRSNorth           &NGC185IRSCenter           &NGC185IRSSouth \\
     &           &Intensity                &Intensity                 &Intensity \\
     &\mum       &$10^{-10}$ W/m$^2$/sr    &$10^{-10}$ W/m$^2$/sr     &$10^{-10}$ W/m$^2$/sr \\
\tableline
H$_2$ S(6)    & 6.10  & 4.4 $\pm$ 1.2  & 2.6 $\pm$ 0.3  & 2.5 $\pm$ 0.4 \\
H$_2$ S(5)    & 6.91  &\nodata         & 3.7 $\pm$ 0.4  & 9.3 $\pm$ 0.4 \\
$[$Ar~II$]$   & 6.98  &\nodata         & 5.9 $\pm$ 0.3  &\nodata        \\
H$_2$ S(4)    & 8.02  &27.3 $\pm$ 4.0\tablenotemark{a}  & 2.0 $\pm$ 0.3  &3.1 $\pm$ 0.5 \\
$[$Ar~III$]$  & 8.99  & 7.4 $\pm$ 0.6  & 4.7 $\pm$ 0.1  &\nodata        \\
H$_2$ S(3)    & 9.66  & 4.7 $\pm$ 0.5  & 2.7 $\pm$ 0.1  & 0.3 $\pm$ 0.1 \\
$[$S~IV$]$    &10.51  & 3.1 $\pm$ 0.5  & 2.5 $\pm$ 0.2  & 2.1 $\pm$ 0.1 \\
H$_2$ S(2)    &12.27  & 2.8 $\pm$ 0.6  & 2.4 $\pm$ 0.3  & 3.8 $\pm$ 1.0 \\
$[$Ne~II$]$   &12.81  & 2.9 $\pm$ 0.5  & 2.3 $\pm$ 0.3  & 5.3 $\pm$ 0.5 \\
$[$Ne~III$]$  &15.55  & 5.4 $\pm$ 0.6  &11.4 $\pm$ 0.3  & 4.7 $\pm$ 0.2 \\
H$_2$ S(1)    &17.03  & 7.4 $\pm$ 1.3  &31.3 $\pm$ 3.0\tablenotemark{a}  &20.7 $\pm$ 2.0\tablenotemark{a} \\
$[$S~III$]$   &18.71  & 4.7 $\pm$ 0.4  & 7.3 $\pm$ 0.2  & 5.6 $\pm$ 0.2 \\
$[$O~IV$]$    &25.91  & 3.6 $\pm$ 0.7  &11.2 $\pm$ 0.7  & 6.5 $\pm$ 0.4 \\
$[$Fe~II$]$   &25.99  & 3.3 $\pm$ 0.6  & 4.6 $\pm$ 0.6  & 5.9 $\pm$ 0.5 \\
H$_2$ S(0)    &28.22  & 3.6 $\pm$ 0.4  & 8.7 $\pm$ 0.4  & 8.6 $\pm$ 0.6 \\
$[$S~III$]$   & 33.48 &\nodata         & 4.4 $\pm$ 0.5  & 2.8 $\pm$ 0.4 \\
$[$Si~II$]$   &34.81  & 4.4 $\pm$ 0.6  &12.6 $\pm$ 4.0\tablenotemark{a}  & 9.0 $\pm$ 3.0\tablenotemark{a} \\
\tableline
\end{tabular}
\tablenotetext{a}{Large uncertainty on the fitted rest wavelength.}
\end{center}
\end{table}

\newpage
\begin{table}[h!]
\centerline{TABLE 11}
\centerline{PAH Emission Features in NGC~185 (Corrected for Extinction) \label{tbl-11}}
\vspace{0.2cm}
\footnotesize
\begin{center}
\begin{tabular}{ccccc}
\tableline\tableline
Feature &Width\tablenotemark{a}   &NGC185IRSNorth           &NGC185IRSCenter           &NGC185IRSSouth \\
        &                         &Intensity                &Intensity                 &Intensity \\
\mum    &\mum                     &$10^{-9}$ W/m$^2$/sr    &$10^{-9}$ W/m$^2$/sr    &$10^{-9}$ W/m$^2$/sr \\
\tableline
 6.2  &0.10  &11.7 $\pm$ 0.7  & 4.8 $\pm$ 0.5  &14.2 $\pm$ 1.0 \\
 7.7  &0.60  &23.8 $\pm$ 0.8  & 7.6 $\pm$ 1.0  &35.3 $\pm$ 2.2 \\
 8.3  &0.10  & 1.9 $\pm$ 0.8  &\nodata         & 3.3 $\pm$ 0.8 \\
 8.6  &0.10  & 1.4 $\pm$ 0.7  &\nodata         & 5.8 $\pm$ 1.0 \\
11.3  &0.20  &13.9 $\pm$ 1.2  &11.8 $\pm$ 1.0  & 8.6 $\pm$ 1.5 \\
12.0  &0.20  & 3.8 $\pm$ 0.4  & 3.9 $\pm$ 0.7  & 2.9 $\pm$ 1.7 \\
12.6  &0.10  & 6.6 $\pm$ 1.4  & 5.0 $\pm$ 0.3  & 6.1 $\pm$ 0.3 \\
13.6  &0.20  &\nodata        & 1.1 $\pm$ 0.3  &\nodata         \\
14.2  &0.10  &\nodata        & 0.9 $\pm$ 0.3  &\nodata         \\
16.4  &0.10  & 0.9 $\pm$ 0.2  & 2.3 $\pm$ 0.4  & 2.4 $\pm$ 0.4 \\
17.0  &1.50  &16.3 $\pm$ 0.4  &10.0 $\pm$ 0.7  & 5.4 $\pm$ 0.5 \\
17.4  &0.10  & 0.2 $\pm$ 0.1  & 1.7 $\pm$ 0.3  & 1.4 $\pm$ 0.2 \\
\tableline
\end{tabular}
\tablenotetext{a}{Width of band of this feature.}
\end{center}
\end{table}

\end{document}